\documentclass[apj]{emulateapj}

\usepackage{natbib}
\usepackage{graphicx}
\bibpunct{(}{)}{;}{a}{}{,}

\newcommand\lsb{{\rm QBCD}}
\newcommand\bcd{{\rm BCD}}

\epsscale{0.5}

\begin{document}

   \title{Search for Blue Compact Dwarf Galaxies During Quiescence}

\author{
	J. S\'anchez Almeida\altaffilmark{1},  
	C. Mu\~noz-Tu\~n\'on\altaffilmark{1},  
	R. Amor\'\i n\altaffilmark{1}, 
	J. A. Aguerri\altaffilmark{1},  \\
	R. S\'anchez-Janssen\altaffilmark{1}, 
	and 
	G. Tenorio-Tagle\altaffilmark{2}
          }
\altaffiltext{1}{Instituto de Astrof\'\i sica de Canarias, E-38205 La Laguna, Tenerife, Spain}
\altaffiltext{2}{Instituto Nacional de Astrof\'\i sica \'Optica y 
	Electr\'onica, AP 51, 72000 Puebla, Mexico}
 
            \email{jos@iac.es, cmt@iac.es, ramorin@iac.es,  
	        jalfonso@iac.es,
	        ruben@iac.es,
		gtt@inaoep.mx}

\begin{abstract}
Blue Compact Dwarf (\bcd) galaxies are metal poor systems going 
through a major
starburst that cannot last for long.
We have identified galaxies which may be 
BCDs during quiescence (\lsb ), i.e.,  before the 
characteristic
starburst sets in or
when it has faded away.
These \lsb\ galaxies are assumed to be like the \bcd\  host 
galaxies.
The SDSS/DR6 database provides $\sim$21500 \lsb\ candidates.
We also select from SDSS/DR6 a complete 
sample of BCD galaxies to serve as reference.
The properties of these two galaxy sets have been
computed and compared.
The QBCD candidates are thirty times more abundant than the BCDs, with
	their luminosity functions being very similar except for the scaling
	factor, and the expected luminosity 
	dimming
	associated with
	the end of the starburst. 
	QBCDs are redder than BCDs, and they have larger H~{\sc ii} 
	region based oxygen abundance.
	\lsb s also have lower surface brightness.
	The BCD candidates turn out to be the QBCD candidates with the largest
	specific star formation rate (actually, with the largest $H_{\alpha}$ 
	equivalent width). One out of each three dwarf galaxies in the 
	local universe may be a QBCD. 
The properties of the selected BCDs and QBCDs are consistent with
	a single sequence in galactic evolution, with the quiescent 
	phase lasting thirty times longer than the starburst phase. 
	The resulting time-averaged star formation rate is low 
	enough to allow 
	this cadence of \bcd\ -- \lsb\
	 phases during the Hubble time.
\end{abstract}



\shorttitle{BCD galaxies during quiescence}
\shortauthors{S\'anchez Almeida et al.}

\section{Introduction}\label{introduction}

Blue Compact Dwarf (BCD) galaxies\footnote{
Also referred to as  H~{\sc ii} galaxies or 
blue amorphous galaxies; see, e.g., \citet{kun00}, \S~4.4.
}
are metal poor systems undergoing 
vigorous star formation \citep[e.g.,][]{thu91,gil03}. 
With record-breaking low metallicities
among galaxies \citep[e.g.,][]{kun00}, 
their 
observed  colors and spectra 
indicate 
essentially newborn
starbursts, with mean ages 
of a few  My \citep[e.g.,][]{mas99,thu91}. 
This combination of factors (chemically unevolved systems, with
oversized starbursts that cannot last for long) 
lead to conjecturing that they are pristine galaxies 
undergoing star formation for the very first time \citep{sar70}. 
This original view has now been reformulated
so that \bcd\ galaxies are chemically primitive 
objects which we come across during short intense bursts. 
The starburst phases are 
interleaved by long periods of quiescence \citep{sea73}.
The change is based on several observational evidences.
Most BCDs galaxies are known to have a red low surface brightness component 
\citep[e.g., ][and references therein]{cao05}, which should exist before 
the starburst, and which  should survive the BCD phase. 
We are also allowed to resolve some individual stars in the nearest 
BCD galaxies, and the presence of RGB stars indicates an
underlying stellar population much older than the lifetime of a 
starburst \citep[e.g.,][]{alo07,cor07}.

Even if they are not pristine, BCDs are galaxies with the lowest 
metallicities and may therefore
be showing 
first stages in star formation from
primordial gas. 
It is not yet clear
which objects
eventually change to glow as  \bcd\
galaxies 
(i.e., which 
galaxies are BCDs during quiescence).
Identifying them seems to be a 
necessary first step
to answer the question of why a galaxy experiences a BCD 
phase. What is the nature of the host galaxy 
in which a
BCD starburst takes place? 
Is it a normal dwarf galaxy?
What is triggering the starburst? 
What is left after the starburst?
Evolutionary connections between dwarf elliptical galaxies, 
dwarf irregular galaxies, 
low surface brightness galaxies,
and BCDs   
have been both proposed and 
criticized in the literature 
\citep[e.g.,][]{sea72,sil87,dav88,pap96,tel97,gil05}.
Here we take a new approach to investigate the
linage of the \bcd s.
Rather than directly comparing 
the properties of the BCDs (or their host galaxies)
with properties of known galaxy types, 
we attempt a blind search for galaxies that look like
precursors or left-overs of the BCD phase, paying
minor attention to the galaxy class.
%
We have tried
to find field galaxies with the properties
of the BCD host, but without the BCD starburst.
Isolated galaxies are preferred to minimize the role of
mergers and harassment on the galactic 
properties and evolution.
These galaxies
 will be denoted in the paper as Quiescent \bcd\
galaxies or \lsb  . 
If there are enough such galaxies, the BCD may be 
just a particularly conspicuous phase in the life 
of otherwise mean dwarf galaxies. If there are no
such galaxies, we should conclude that the 
BCD phase ends up with the host, 
which is physically unlikely.
Aiming at shedding light on the \bcd\ evolution, 
we have tried to answer two specific questions:
(1) are there galaxies like the
\bcd\  hosts without the conspicuous starburst 
observed in \bcd s? 
(2) If the answer were yes (as it turns out to 
be), what are the physical properties
of these (putative) BCD galaxies during
quiescence?

The different sections of this paper describe
individual steps carried out to answer the 
two questions posed above. The search for
\lsb\ galaxies like the BCD host galaxies is described
in \S~\ref{lsbstuff}. Luminosities, surface brightnesses, 
and colors characteristic of \bcd\ host galaxies are
taken from \citet{amo07,amo07b}. We search the SDSS/DR6 database
for candidates, which is 
ideal for the kind of exploratory statistical study
we aim at. 
The sample of \lsb\ galaxies has to be 
interpreted in terms of a sample of 
BCD galaxies. In  order to minimize systematic errors,
the control sample of BCDs is also derived 
from SDSS/DR6 using the same techniques employed
for \lsb\ selection (\S~\ref{select_bcd}). 
We compare the Luminosity Functions
(LFs) of the two samples
in \S~\ref{lf}, which requires computing how
the BCD galaxy luminosity changes to become
a \lsb\ galaxy when the starburst fades away
(\S\ref{bcd2lsb}).
Since the SDSS catalog is magnitude
limited, the properties that we derive are biased
towards the brightest galaxies (Malmquist bias).
This effect is corrected  
to derive the intrinsic properties 
of the two sets (\S~\ref{malmquist}).
The implications of our findings are discussed
in \S~\ref{discusion} and \S~\ref{conclusions}.
We have developed our own software to compute
LFs. For the sake of clarity 
and future reference, 
details are given in 
App.~\ref{appa}.
A Hubble constant $H_0=70$~km~s$^{-1}$~Mpc$^{-1}$
is used throughout the paper.

\section{Data sets}\label{data_set}

The search has been carried out using the 
Sloan Digital Sky Survey (SDSS) dataset,
which is both convenient and powerful.
We use the latest data release, DR6, whose 
spectroscopic view  
covers  7425~deg$^2$ and contains
$\sim 7.9\times 10^{5}$ galaxies \citep{ade08}.
The main characteristics of the SDSS  are described in the 
extensive paper by \citet{sto02}, but they are
also gathered in the 
comprehensive searchable  SDSS
website\footnote{{\tt http://www.sdss.org/dr6}}.
We access the database through the so-called
CasJob entry, which allows direct and flexible 
SQL searches, such as those
needed to select only isolated galaxies. 
Absolute magnitudes are computed from relative 
magnitudes and redshifts.  
All magnitudes have been 
corrected for galactic extinction \citep{sch98}.
No $K-$correction has been applied since 
we are dealing with nearby galaxies 
with redshifts $< 0.05$, and so, with 
Hubble flow induced bandshifts
much smaller than the relative bandwidth of the color 
filters \citep{fuk96}.
In a consistency test using SDSS spectra 
representative of QBCDs and 
BCDs, we have estimated that the $K-$corrections
are smaller than 0.01 magnitudes (QBCDs) and
0.1 magnitudes (BCDs).

\subsection{Selection of galaxies like the hosts of BCDs}
	\label{lsbstuff}

\begin{figure}
\includegraphics[width=0.5\textwidth]{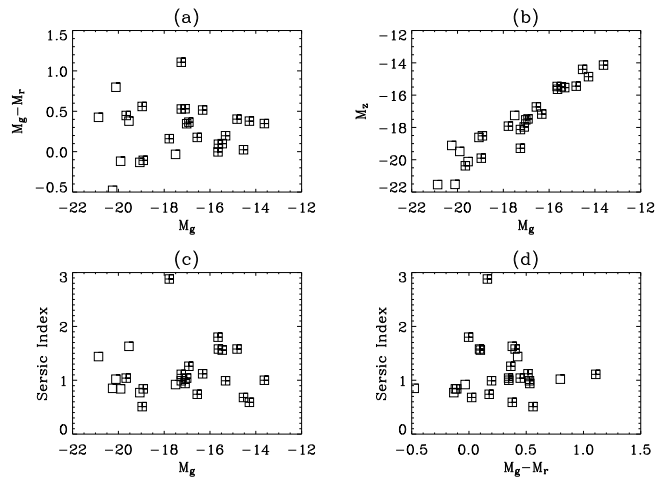}
\caption{
Scatter plots with magnitudes, colors and concentration indexes 
characterizing the  hosts of BCD galaxies
according to \citet{amo07,amo07b}. (a) Color vs absolute magnitude. (b)
Redest SDSS magnitude $z$ (9097~\AA ) vs visible magnitude $g$ 
(4825~\AA ).
(c) Sersic index ($\equiv$ concentration) vs magnitude. 
(d) Sersic index vs color.
The plus symbols inserted into the boxes mark the
gold subsample, i.e., those galaxies 
which are believed to be characterized best.
}
\label{lsb_data}
\end{figure}

The properties of the \lsb\  galaxies
have been taken from  the sample of 28 BCD galaxies 
in \citet{cai01c},
whose hosts have been characterized by \citet{amo07,amo07b}. 
They represent a large set of host galaxy properties 
consistently derived from 2-dimensional fits.
\citet{amo07,amo07b} fit Sersic 
profiles\footnote{The magnitude $m$ varies with the 
radial distance from the center of the galaxy $\rho$ as 
$m(\rho)=m(0)-b_n[\rho/R_e]^{1/n}$,
with $b_n$ chosen so that half of the galaxy light is enclosed
within $\rho < R_e$ \citep[see, e.g.,][]{cio91}. The
so-called Sersic index $n$ controls the shape of the profile.
} 
to the outskirts of each BCD galaxy
once the bright central blue component has been masked out. 
From fits in various bandpasses, they derive 
luminosities, colors and Sersic indexes.
We use them to guide the SDSS search. 
(The impact on the selection of using this particular
characterization 
rather than other alternatives is discussed in \S~\ref{characterization}.)
The actual properties of these
\bcd\ host galaxies are shown in Fig.~\ref{lsb_data}. 
The SDSS photometric colors $u\,g\,r\,i\,z$ are not the standard 
$UBVRI$ system used by \citeauthor{amo07b}
Therefore, in order to produce Fig.~\ref{lsb_data},
the original de-reddened colors have been translated
to SDSS magnitudes using the recipe in \citet[][Table 7]{smi02}.
This photometric transformation is suited for stars but, 
according to the SDSS website, the equations also provide 
reliable results for galaxies without strong emission lines, 
as we expect the \bcd\ hosts to be.

Note that the absolute 
magnitude, the color, and the Sersic
index of each \bcd\ host 
do not seem to be correlated with each other
(cf. Figs.~\ref{lsb_data}a, \ref{lsb_data}c, and  \ref{lsb_data}d).
This fact allows us to assume that the magnitude, color and
concentration of each host galaxy are independent, and we will 
look for  
galaxies whose properties span the full range of observed values.

\begin{figure}
   \centering
   \includegraphics[width=0.5\textwidth]{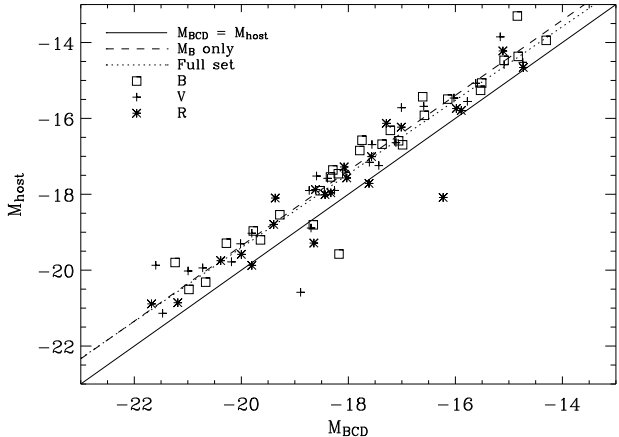}
     \caption{Absolute magnitude of the low surface brightness
	 host  of the BCD
	vs magnitude of the BCD, including the host.
	Each symbol corresponds to a Johnson's bandpass as labeled 
	in the inset. There is a tight correlation between the
	magnitude of the \bcd\ and the magnitude of the
	 host galaxy. A linear least squares fit based on  $B$
	magnitudes yields a slope of $0.991\pm 0.054$, and an offset of
	$0.61\pm 0.11$ (the dashed line).
	Within error bars, the results are identical when using the full
	data  set for fitting (the dotted line),
	indicating a relationship independent of the color.
	The  gold set defined by \citet{amo07b} also yields
	the same result (not shown). 
	} 
         \label{mag_vs_mag}
   \end{figure}

Before we can proceed, 
there is an important observational property of the \bcd\ hosts 
that needs to be 
mentioned. (For a detailled discussion, see \citealt{amo07b}.)
BCD galaxies and their host galaxies tend to have similar 
luminosities, so that the brighter the BCD galaxy the brighter 
the host \citep[see][]{pap96,mar99b}.
Actually, there seems to be a simple relationship between the 
magnitude of the host, $M_{\rm host}$,
and the magnitude of the BCD galaxy, $M_{\rm BCD}$, 
\begin{equation}
M_{\rm host}\simeq 0.5+M_{\rm BCD}.
\label{law}
\end{equation}
Figure~\ref{mag_vs_mag} presents the scatter plot between the
observed magnitude of the BCDs analyzed by \citet[][]{amo07b},
and the magnitude of their host galaxy.
There is a linear relationship between
the two observed magnitudes as parameterized by equation~(\ref{law})
-- the slope is very close to
one, and the scatter remains rather small (less than 0.5 mag).
The relationship is independent of the color and,
therefore, it will remain like equation~(\ref{law}) for 
the SDSS photometric system.

%
%
%
%

\begin{deluxetable}{cll} 
\tablecolumns{3} 
\tablewidth{0pt} 
\tablecaption{Criteria used to select \bcd\ host-like galaxies (i.e.,
\lsb s) from the SDSS/RD6 database.
\label{criteria_lsb}
}
\tablehead{ 
\colhead{\#} & \colhead{Criterion}     &  \colhead{Implementation} 
}
\startdata 
(1) &    Colors & $-0.5 \le M_g-M_r \leq 1.1$ mag\\
(2)& Concentration indexes  & $ 1.8 < R_{90}/R_{50}< 3.1$\\
(3)& Magnitudes & $-18.5 < M_g <  -13.5$  mag\\
(4)&Be isolated& no bright galaxy within 
	10 $R_{50}$\tablenotemark{a}\\
(5)&Get rid of proper motion& redshift $> 0.003$\\
&induced redshifts&\\
\enddata 
\tablenotetext{a}{Bright means brighter than 3 magnitudes fainter
	than the selected galaxy.}
\end{deluxetable}

Table~\ref{criteria_lsb} lists the actual constraints employed 
in our SDSS/DR6 search for 
\lsb\ galaxies like the BCD host galaxies. 
The range of colors,  
\begin{equation}
-0.5\le M_g-M_r \le 1.1,
\end{equation}
has been taken directly from Fig.~\ref{lsb_data}a.
The range  of absolute magnitudes, 
\begin{equation}
-18.5\le M_g \le -13.5,\label{abs_mag_cons}
\label{noname}
\end{equation}
requires a more elaborated explanation.
The upper bound corresponds to the magnitude of the faintest
\bcd\ host (Fig.~\ref{lsb_data}a). The lower limit,
however, is inherited from the lower limit 
of the BCD galaxies selected 
in \S~\ref{select_bcd}. This is a control set used for 
reference and, according to arguments to be given 
in \S~\ref{select_bcd}, 
SDSS BCD galaxy candidates are chosen to be  
fainter than $M_g\simeq -19$. Therefore, for
consistency with equation~(\ref{law}),
we impose the lower limit given
in equation~(\ref{noname}).
There is an additional detail concerning the magnitudes in use. 
\citet{amo07,amo07b} integrate the Sersic profile to infinity 
to estimate magnitudes, whereas the SDSS catalog 
provides Petrosian magnitudes, where the 
galactic light is integrated only up to a certain
distance from the galactic center \citep[see][]{sto02}.
\citet{gra05} show how the difference between the Petrosian
magnitude
and the total magnitude differs by less than 0.2~magnitudes 
for Sersic indexes $n < 4$. The hosts have Sersic indexes smaller
than this limit (Fig.~\ref{lsb_data}c) and, therefore, we 
neglect 
the difference.

The brightness profile of the \bcd\ host is parameterized in
\citet{amo07,amo07b} by means
of the Sersic index $n$. According to Fig.~\ref{lsb_data}c,
\begin{equation}
0.5 < n < 2.9. 
\label{index_cons}
\end{equation}
This constraint is translated into a 
limit on the so-called concentration index $R_{90}/R_{50}$,
which can be readily obtained from the SDSS catalog since
it provides the radii containing 50\% of Petrosian flux,
$R_{50}$, and 90\% of the Petrosian flux, $R_{90}$.
According to 
\citet[][Table 1]{gra05},
the limits given in equation~(\ref{index_cons})
imply,
\begin{equation}
 1.8 < R_{90}/R_{50} < 3.1,
\end{equation}
which is the constraint used in our search
(Table~\ref{criteria_lsb}, 2nd row).
This constraint has been applied both to $g$ 
and to $r$ magnitudes.

As explained in the introduction,
we seek isolated galaxies
to minimize the effect of mergers and harassment
on the galaxy properties. 
The criterion
for finding
isolated galaxies is inspired by 
the work of \citet{all05}.
We  stipulate that 
the selected galaxies have no bright companion 
within $10~R_{50}$. (Note that
$R_{50}$ is approximately the effective radius, i.e.,
the radius containing half of the galaxy luminous flux.) 
Companions are neglected if they
are at least three magnitudes fainter than the
candidate. These two constraints
are set in the color filter $g$ (see Table~\ref{criteria_lsb}, 
4th row).

Since absolute magnitudes are computed from redshifts,
we ask the redshifts to be large enough
to minimize the proper-motion induced Doppler shifts. 
The threshold redshift
corresponds to a distance of 13~Mpc and a velocity of 900 km~s$^{-1}$
(see Table~\ref{criteria_lsb}, 5th row).

\begin{figure*}

\includegraphics[width=0.24\textwidth]{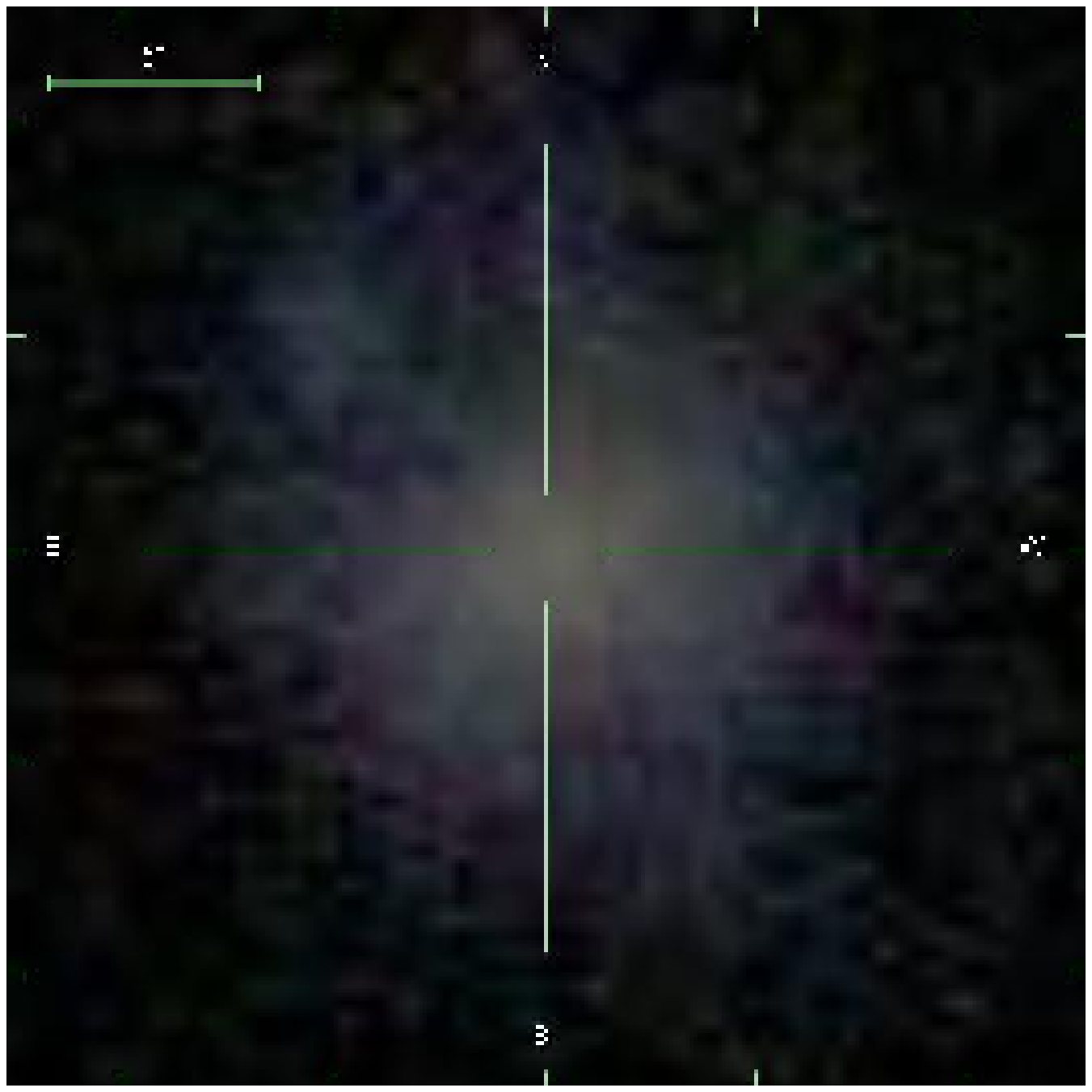}
\includegraphics[width=0.24\textwidth]{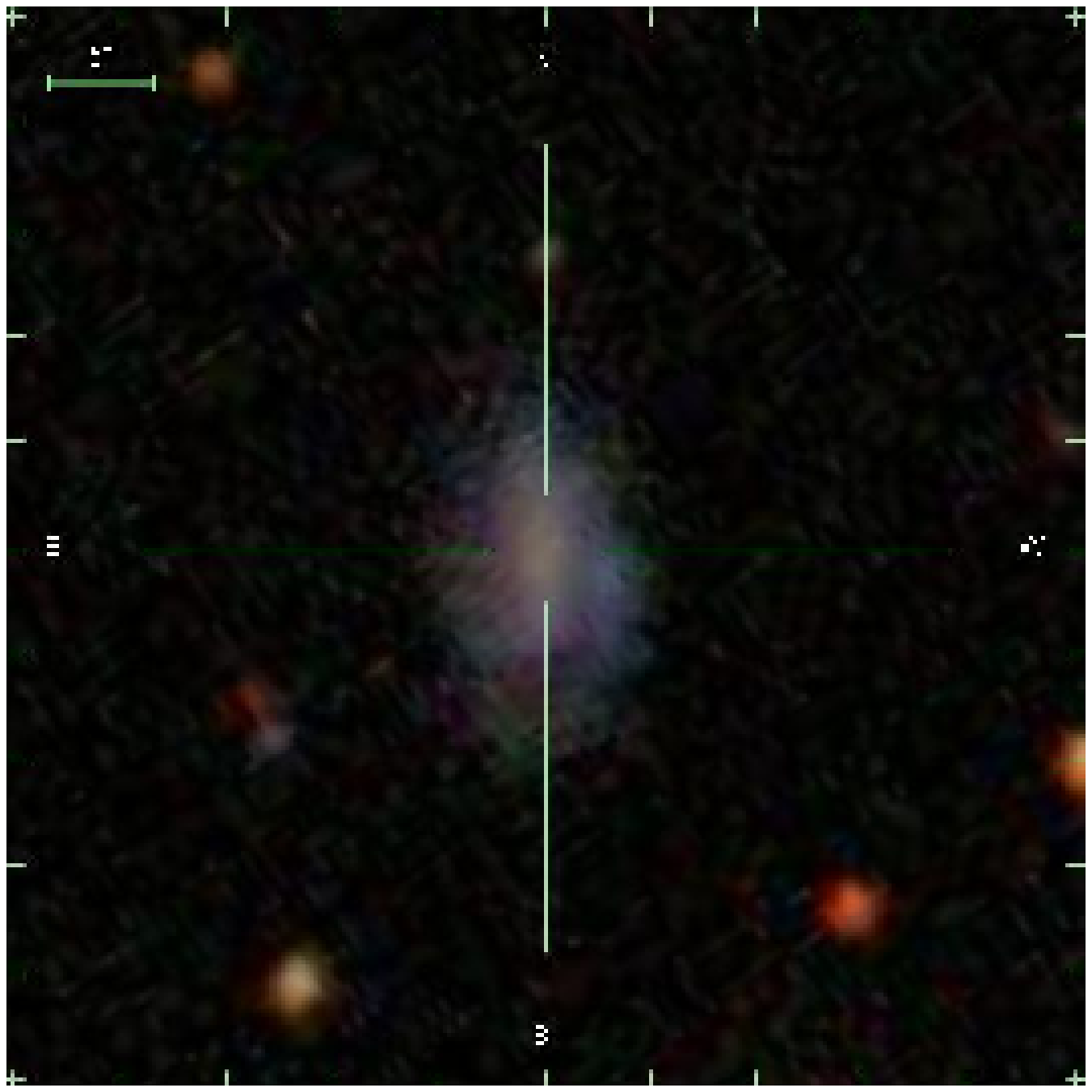}
\includegraphics[width=0.24\textwidth]{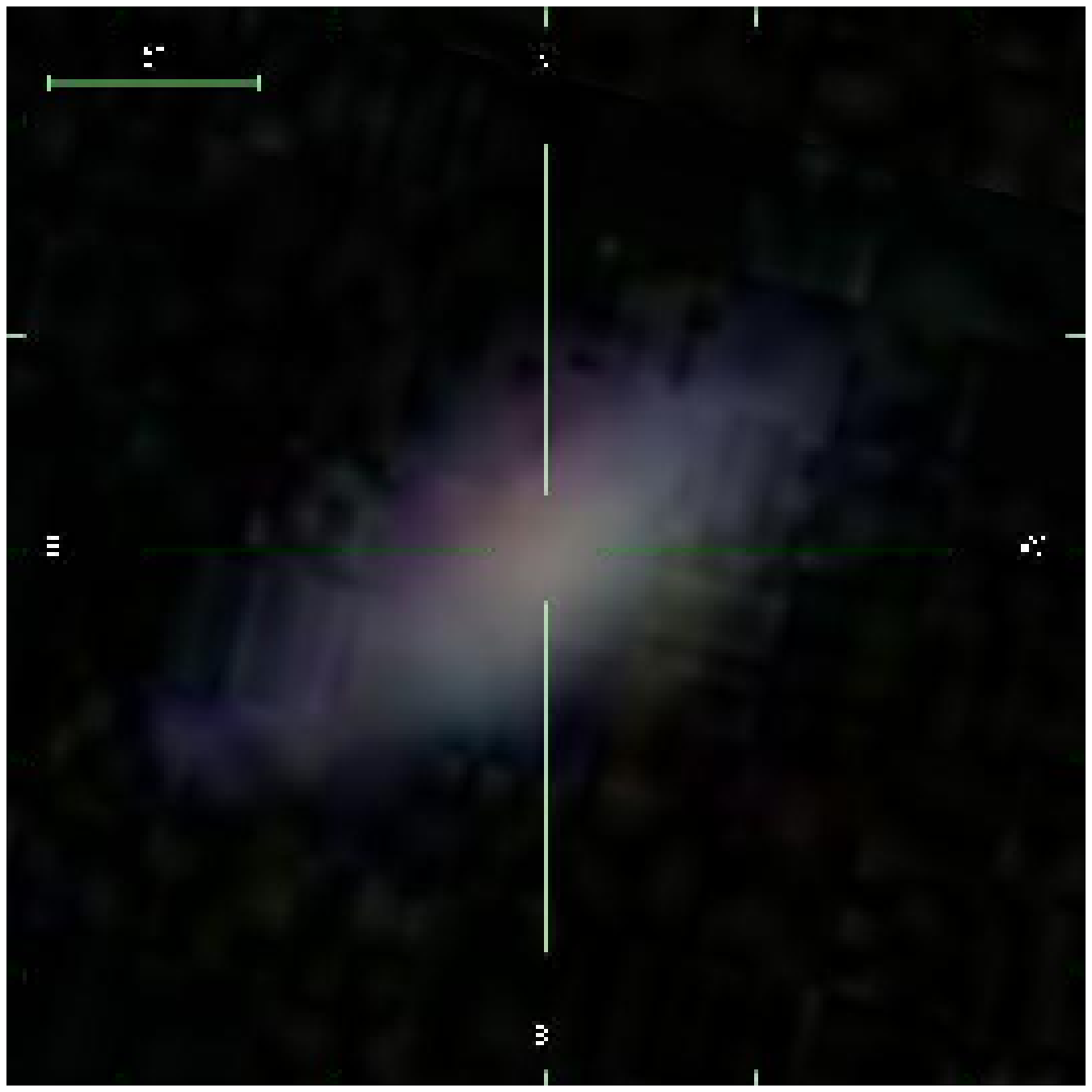}
\includegraphics[width=0.24\textwidth]{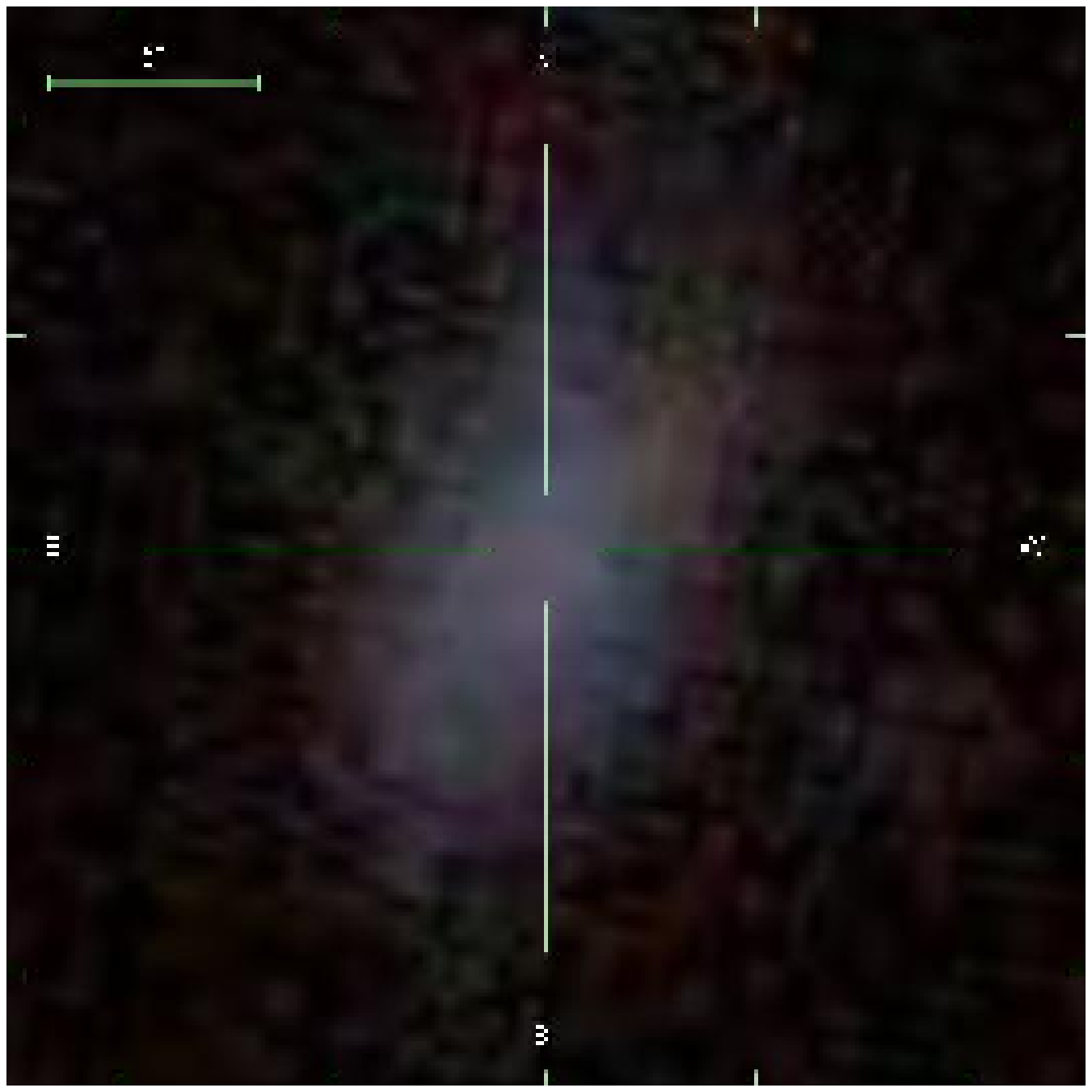}

\includegraphics[width=0.24\textwidth]{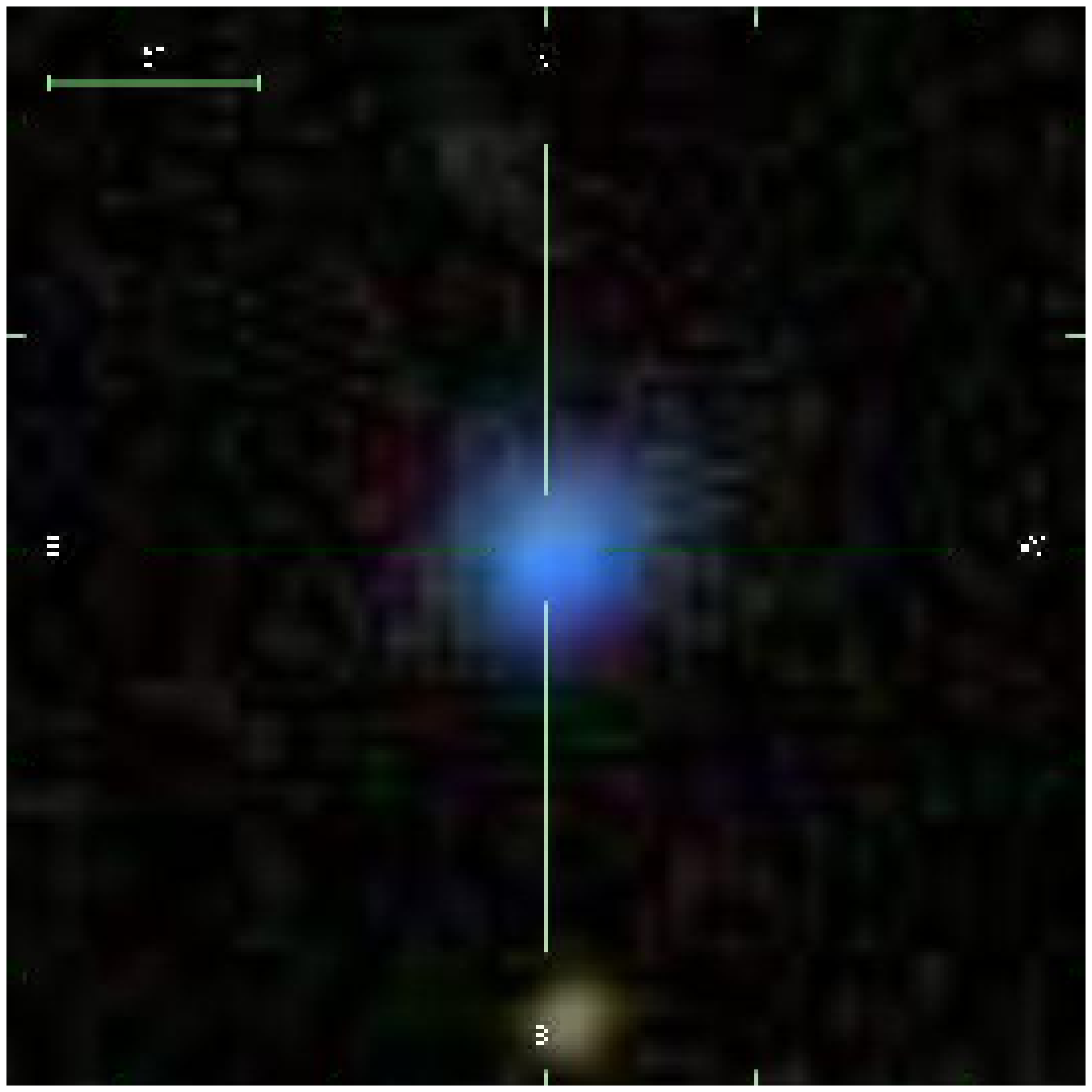}
\includegraphics[width=0.24\textwidth]{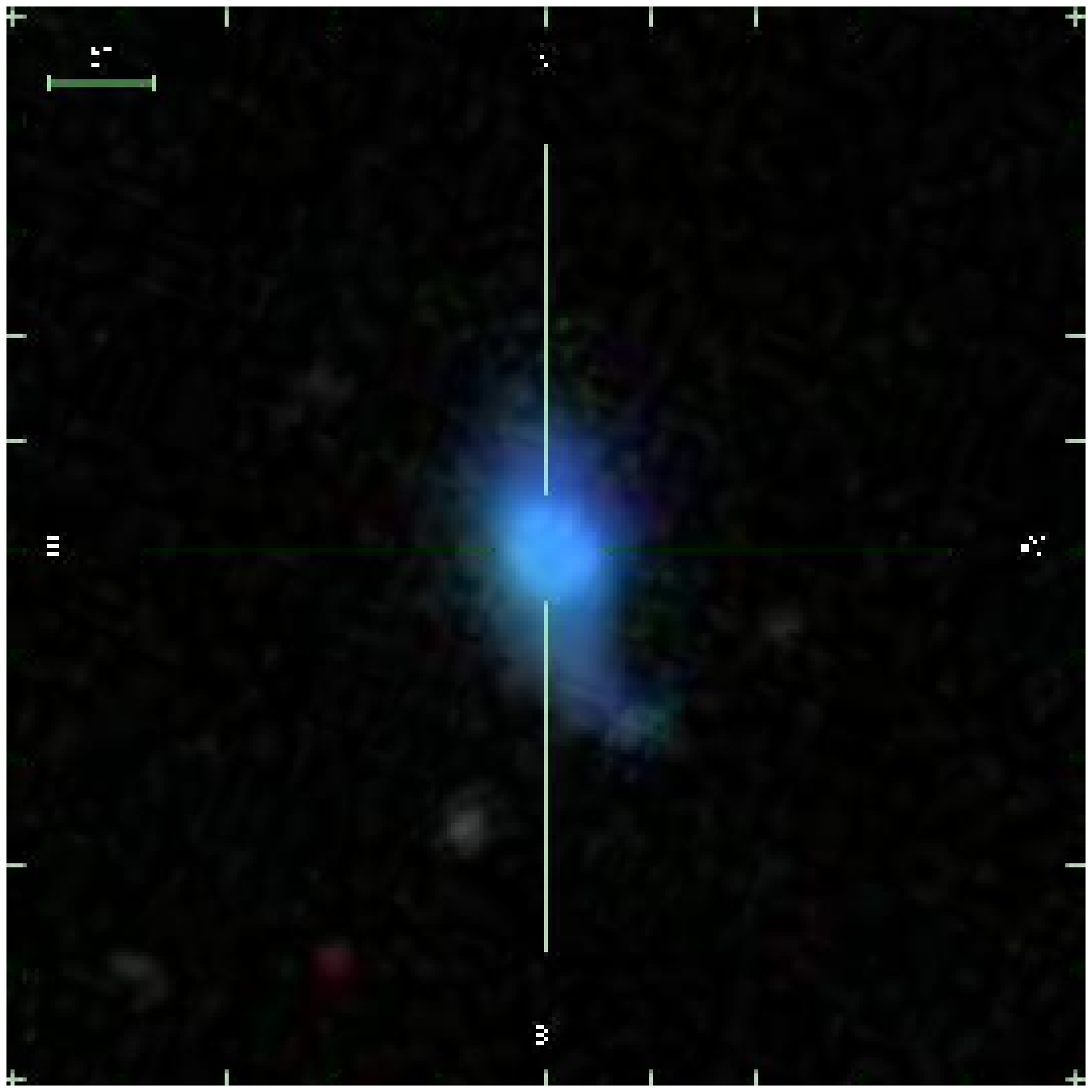}
\includegraphics[width=0.24\textwidth]{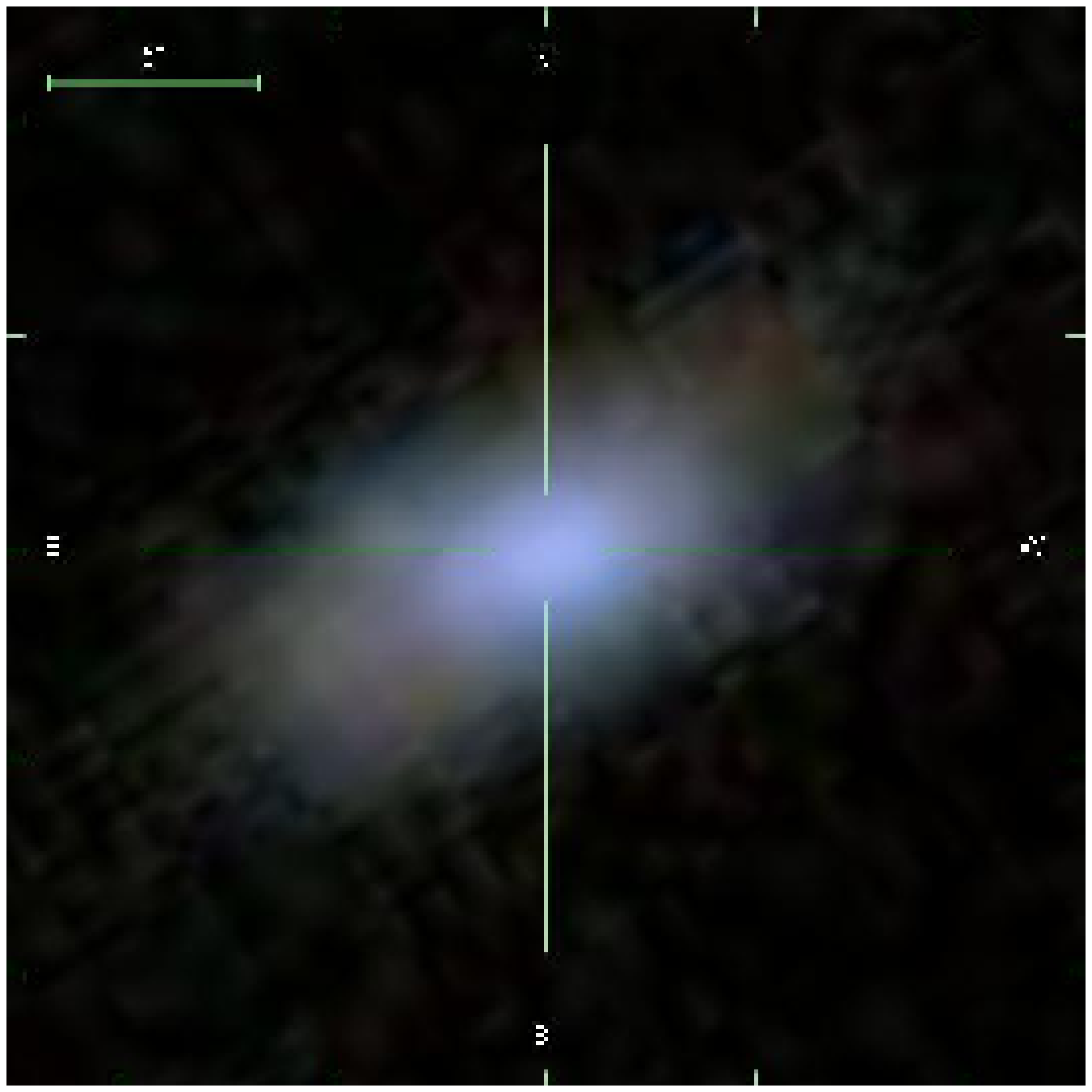}
\includegraphics[width=0.24\textwidth]{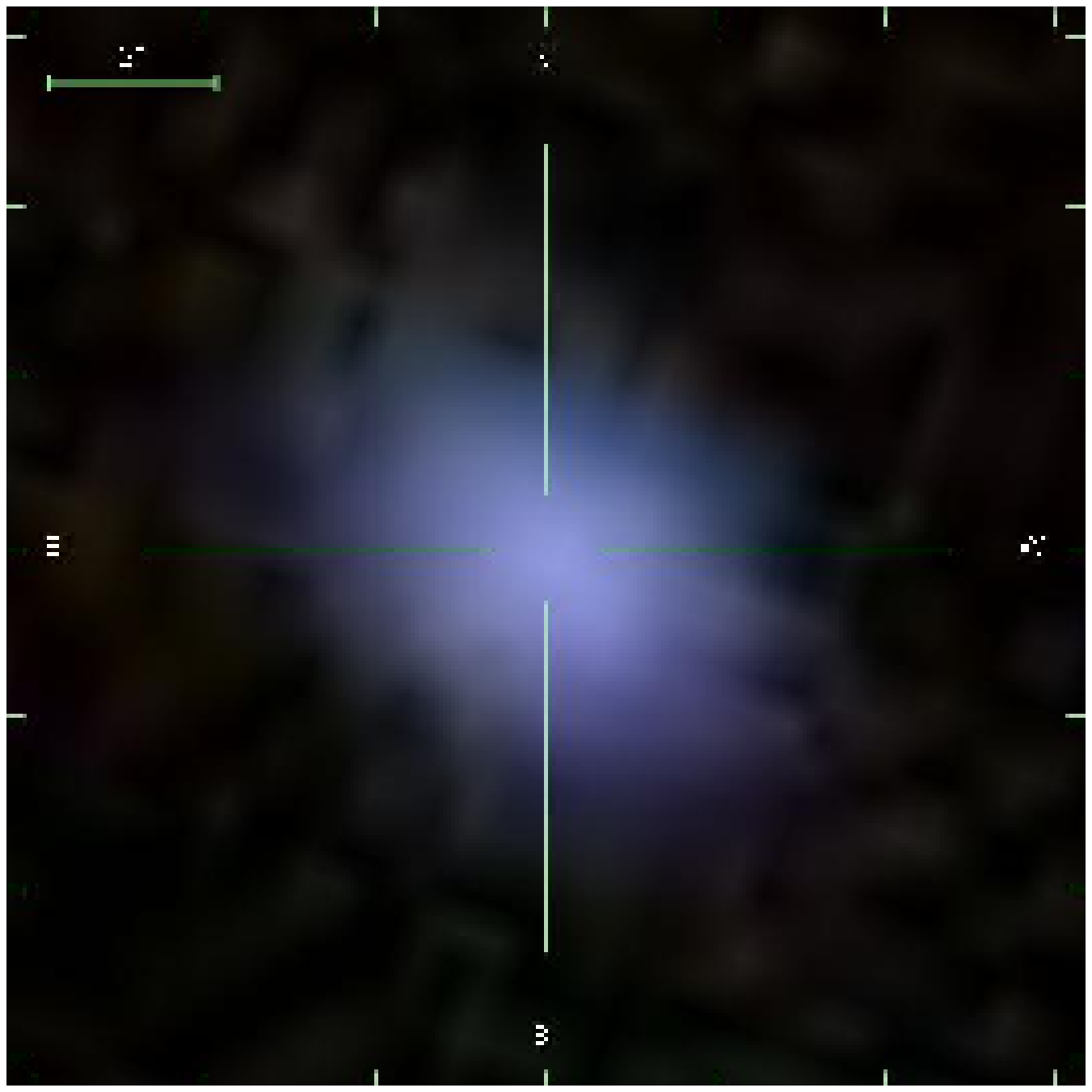}
\caption{
Set of randomly chosen images of QBCD candidates (top) and BCD candidates (bottom). 
We are using the color look-up table of SDSS.
Each galaxy includes a horizontal scale  corresponding to 5\arcsec . The crosses centered in the
galaxies point 
out north, south, east and west directions.
}
\label{image_sample}
\end{figure*}

\begin{figure*}
\includegraphics[width=0.7\textwidth,angle=90]{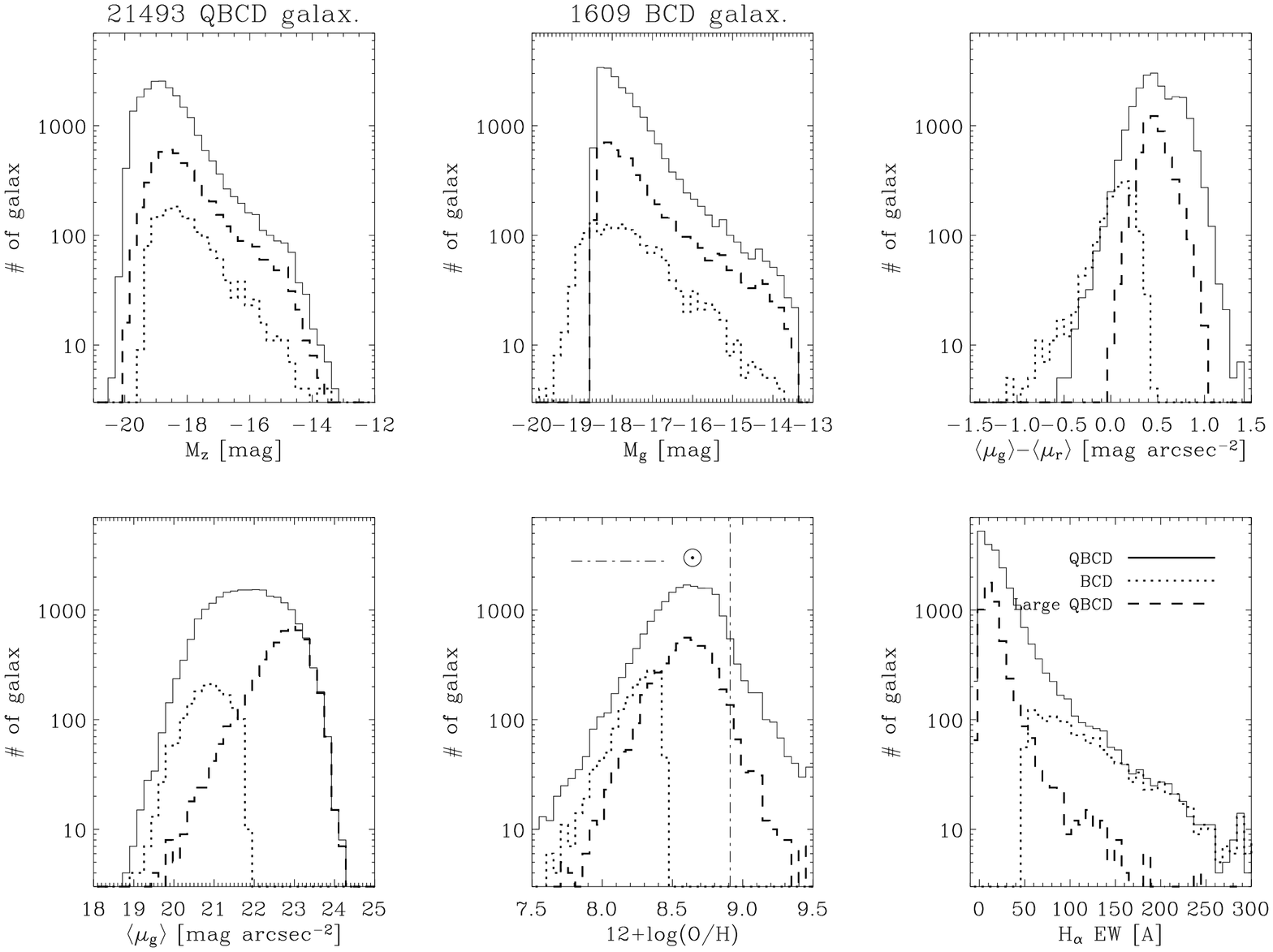}
\caption{
Histograms of the observed magnitudes and colors corresponding to
our selection of \lsb\ galaxies (the solid lines), and BCD candidates
(the dotted lines). We also consider a sub-set of 
\lsb\ galaxies with large apparent size ($R_{50} > 4$\arcsec).
Their properties are shown using dashed lines. The labels
in the ordinate axes specify the parameter that is represented --
from top to bottom and left to right, absolute $z$ magnitude
$M_z$, absolute $g$ magnitude $M_g$,
surface brightness color 
$\langle\mu_g\rangle-\langle\mu_r\rangle$,
surface brightness $\langle\mu_g\rangle$,
oxygen-to-hydrogen abundance in the usual scale $12+\log({\rm O/H})$,
and $H_\alpha$ equivalent width.
The histogram of metallicities also includes the solar value for reference
($12+\log[{\rm O/H}]_{\odot}\simeq 8.91$;  \citealt{kun00}).
}\label{myBCD}
\end{figure*}

Applying the criteria in Table~\ref{criteria_lsb} to the SDSS/DR6, 
one retrieves 21493 galaxies. 
Their mean redshift is 0.030, with a standard deviation of 0.014. 
The galaxies 
are illustrated in Fig.~\ref{image_sample}, top row, where we include
color images of four randomly chosen {\lsb}\ candidates.
Figure~\ref{myBCD} shows various
histograms corresponding to the observed properties of 
this set of galaxies
 (the solid lines).
Means, standard deviations and modes corresponding to
these histograms are listed
in Table~\ref{properties}. 
The determination of the concentration indexes of
small galaxies may 
be problematical.
They are based
on determining galaxy sizes and, therefore,
they are affected by seeing \citep[see][]{bla03}.
We select a subset among the \lsb\  candidates
to test whether the main properties of the candidates
depend on the spatial resolution. To be 
on the safe side,
we consider candidates having $R_{50} > 4$\arcsec, i.e.,
larger than the typical SDSS seeing ($\la 1.5\arcsec$). 
The corresponding histograms are also 
shown in Fig.~\ref{myBCD}
(the dashed lines), with the
means, standard deviations and modes  
included in Table~\ref{properties}.
Except for the surface brightness,
there are no systematic differences between the 
full set and the subset of large \lsb\ galaxies.
Moreover, the decrease of surface brightness for
large galaxies has nothing to do with 
poor seeing. It is a bias imposed by our  
lower limit in absolute magnitude (Table~\ref{criteria_lsb}, 3rd
row). Unless the large galaxies are also
low surface brightness, they are much too bright to 
satisfy our selection criterion. In short, 
poor seeing does not seem to bias our selection in any obvious way.
%


\begin{deluxetable*}{cccccccccccccccccl} 
\tabletypesize{\tiny}
\tablecolumns{8} 
\tablewidth{0pc} 
\tablecaption{Statistical properties of various galaxy sets.\label{properties}}
\tablehead{ 
\colhead{Galaxy}    
&\multicolumn{2}{c}{$M_z$}&\colhead{}
&\multicolumn{2}{c}{$M_g$}&\colhead{}
&\multicolumn{2}{c}{$\langle\mu_g\rangle-\langle\mu_r\rangle$}&\colhead{}
&\multicolumn{2}{c}{$\langle\mu_g\rangle$}&\colhead{}
&\multicolumn{2}{c}{12+log({\rm O/H})}&\colhead{}
&\multicolumn{2}{c}{$H_\alpha$ EW\tablenotemark{a}}
\\ 
\colhead{Set}    
&\multicolumn{2}{c}{}&\colhead{}
&\multicolumn{2}{c}{}&\colhead{}
&\multicolumn{2}{c}{[mag~arcsec$^{-2}$]}&\colhead{}
&\multicolumn{2}{c}{[mag~arcsec$^{-2}$]}&\colhead{}
&\multicolumn{2}{c}{}&\colhead{}
&\multicolumn{2}{c}{[\AA]}
\\ 
\cline{2-3} 
\cline{5-6} 
\cline{8-9} 
\cline{11-12} 
\cline{14-15} 
\cline{17-18} 
\\ 
\colhead{} & 
\colhead{av\tablenotemark{b}$\pm$std\tablenotemark{c}}   & \colhead{md\tablenotemark{d}}    
& \colhead{}&
\colhead{av$\pm$std}   & \colhead{md}    & \colhead{}&
\colhead{av$\pm$std}   & \colhead{md}    & \colhead{}&
\colhead{av$\pm$std}   & \colhead{md}    &\colhead{}&
\colhead{av$\pm$std}   & \colhead{md}    &\colhead{}&
\colhead{av$\pm$std}   & \colhead{md}    
}
\startdata 
Observed \lsb &-18.5$\pm$1.1&-18.8&&
-17.6$\pm$0.8&-18.3&&
0.50$\pm$0.24&0.46&&
21.8$\pm$0.88&22.0&&
8.61$\pm$0.30&8.60&&28.1$\pm$49.2& 1.9\\
Large \lsb &-18.0$\pm$1.3&-18.6&&
-17.3$\pm$1.0&-18.1&&
0.47$\pm$0.16&0.46&&
22.7$\pm$0.67&22.9&&
8.61$\pm$0.26&8.60&&
19.9$\pm$35.5& 9.9\\
Observed BCD&-17.9$\pm$1.1&-18.3&&
-17.5$\pm$1.1&-19.0&&
0.04$\pm$0.23&0.15&&
20.8$\pm$0.54&21.0&&
8.24$\pm$0.16&8.35&&
172.$\pm$174.& 58.\\
%
Restored \lsb &-16.2$\pm$1.6&-14.9&&
-15.5$\pm$1.4&-14.2&&
0.37$\pm$0.23&0.38&&
22.2$\pm$0.95&22.6&&
8.43$\pm$0.31&8.40&&
33.1$\pm$50.4& 1.9\\
Restored BCD&-16.2$\pm$1.5&-16.3&&
-15.8$\pm$1.4&-15.1&&
0.04$\pm$0.22&0.15&&
21.0$\pm$0.59&21.7&&
8.12$\pm$0.20&8.14&&
121.$\pm$ 64.& 58.\\
\enddata 
\tablenotetext{a}{$H_\alpha$ equivalent width}
\tablenotetext{b}{Average}
\tablenotetext{c}{Standard deviation}
\tablenotetext{d}{Mode}
\end{deluxetable*} 

\subsection{Selection of BCD galaxies}\label{select_bcd}

The galaxies in  \S~\ref{lsbstuff} are selected to be
quiescent BCDs. It is clear that this conjecture and 
the galaxy properties must be examined in terms of 
the properties of the BCD galaxies. This 
is particularly true to answer the first basic question
of whether the number density of \lsb\ candidates suffices 
to account for the existing BCDs.
Number densities are best characterized as LFs but, 
to the best of our knowledge, there is no BCD LF
in the literature. Moreover, even if such 
a LF existed, 
it would have been produced with a number of (unknown) 
biases different from those involved
in our SDSS selection.
Therefore, 
we found it necessary to extract from SDSS a set 
of BCD galaxies, that is to say, to construct a
reference BCD sample with the biases and problems of 
the \lsb\ galaxies we want to compare to. This section 
describes such selection of BCD candidates.

The general criteria have been taken form \citet{mal05}, but they
are implemented according to \citet{gil03}. 
A BCD galaxy should have the following properties, 
(1) be {\em blue} enough, which
	constrains the colors, 
(2) be {\em compact}, which limits the 
	surface brightness,
(3) be {\em dwarf}, which sets a lower limit to
	the magnitude,
(4) have a {\em large star formation rate} (SFR), which
	implies having enough H~{\sc ii} regions and,
	therefore, enough $H_\alpha$ emission,
(5)  be {\em metal poor}, and
(6)  {\em not to be confused with {\rm AGN}s}.
These general criteria have been specified as 
detailled in Table~\ref{criteria_bcd}. The numerical
values corresponding to
constraints  (1), (2) and (3) have been taken from 
\citet{gil03}, who point out three photometric criteria
for a galaxy to be a BCD,
\begin{displaymath}
\langle\mu_B\rangle-\langle\mu_R\rangle \leq {\rm ~1~mag~arcsec}^{-2},
\end{displaymath}
\begin{equation}
\langle\mu_B\rangle < {\rm ~22~mag~arcsec}^{-2},
\label{caca}
\end{equation}
\begin{displaymath}
M_K > -21 {\rm mag}.
\end{displaymath}
Using the transformation between Johnson's and SDSS
photometric systems in \citet{smi02}, plus equation~(1) in \citet{gil03}, 
these three conditions become the constraints (1), (2) and (3) in
Table~\ref{criteria_bcd}.
The mean surface brightness $\langle\mu\rangle$ has been computed as the 
magnitude of the average luminosity within $R_{50}$, i.e.,
$\langle\mu\rangle=2.5\log(2\pi)+m+5\log R_{50}$, with $m$ the apparent
magnitude \citep[see, e.g.,][\S~2.3]{bla01}.
The transformation between photometric systems
is suited for stars, rather than for galaxies with 
emission lines like the BCDs. However, the 
approximation suffices because the actual limits
in equation~(\ref{caca}) are estimative, and 
the effect of including lines modifies the 
BCD magnitudes by 
0.1 mag or less. This effect has been estimated using 
SDSS spectra representative of BCD galaxies.
%

%
%

\begin{deluxetable}{cll}
\tablecolumns{3} 
\tablewidth{0pt} 
\tablecaption{Criteria used to select BCD galaxies from the SDSS/DR6 database.
\label{criteria_bcd}
}
\tablehead{ 
\colhead{\#} & \colhead{Criterion}     &  \colhead{Implementation} 
}
\startdata 
(1) &    Be blue enough & $\langle\mu_g\rangle-\langle\mu_r\rangle \leq 0.43$ mag arcsec$^{-2}$\\
(2)& Be compact &$\langle\mu_g\rangle < 21.83 - 0.47 (\langle\mu_g\rangle-
	\langle\mu_r\rangle)$ mag arcsec$^{-2}$\\
(3)& Be dwarf& $M_g > -19.12+1.72 (M_g-M_r)$ mag\\
(4)&Having large  SFR& ${H_\alpha}$ Equivalent Width $> 50~$\AA \\
(5)&Be metal poor& $12+\log({\rm O/H}) < 8.43$ ($\equiv$  1/3 $\odot$)\\
(6)&Not to be confused with AGNs & neglect AGN contamination\\
(7)&Be isolated& no bright galaxy within 
	10 $R_{50}$\tablenotemark{a}
	\\
(8)&Get rid of proper motion& redshifts $> 0.003$\\
&induced redshift&\\
\enddata 
\tablenotetext{a}{Bright means brighter than 3 magnitudes fainter
	than the selected galaxy.}
\end{deluxetable} 


The actual implementation of criteria  (4) and (5) are mere 
educated guesses that try not to be too restrictive. For example,
we consider metallicities smaller than  1/3 the solar value 
(Table~\ref{criteria_bcd}, item 5),
when 
\citet[][]{kun00} bound the BCDs metallicities between 1/10 and
1/50.  The (oxygen) metallicity of the galaxies
(O/H) has been
estimated using the so-called N2~method, 
based on the equivalent widths of the 
emission lines $[NII]\lambda6583$ and $H_\alpha$,
\begin{equation}
12+\log({\rm O/H})= 9.12 + 0.73\, {\rm N2},
\end{equation}
with
\begin{displaymath}
{\rm N2}=\log{{[NII]\lambda6583}\over{H_\alpha}};
\end{displaymath}
see \citet{shi05} and references therein.
Note that this metallicity characterizes 
the properties of the H~{\sc ii} regions
excited by the starburst, which represents only
a very small fraction of the galactic gas 
(see \S~\ref{discusion}).
Keeping the appropriate caveats in mind,
the N2~method suffices for the elementary 
estimate we are interested in.

\begin{figure}
\includegraphics[width=0.5\textwidth]{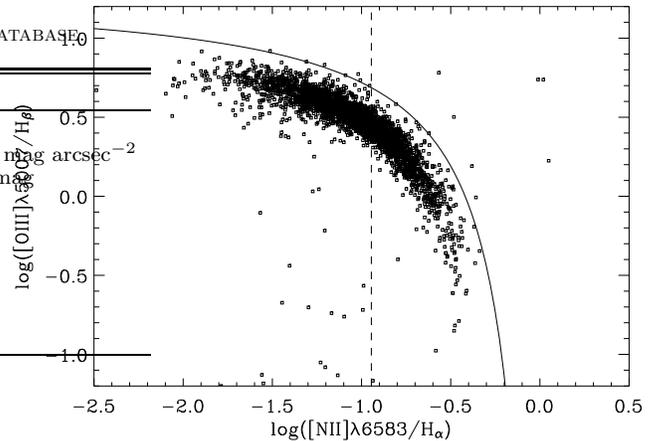}
\caption{
Diagnostic plot used to 
identify AGNs contaminating
the \bcd\ sample. The points above the thick
solid line correspond to AGNs candidates. The vertical dashed
line separates the low O abundance region (to the left) and
the high O region (to the right). We only use low O candidates,
where the AGN contamination is minimum.
The two indexes are defined in the main text.
}
\label{agn}
\end{figure}

	Active Galactic Nuclei (AGNs) can be misidentified 
as star-forming galaxies
since they are high surface brightness blue galaxies
with emission lines. AGNs introduce spurious BCD
candidates. 
In practice, however,  this contamination is 
negligible, and we do not decontaminate 
for the presence 
of AGNs  (item 6 in Table~\ref{criteria_bcd}).
Our approach can be justified using the criterion to be an AGN 
by \citet[][]{kau03}.
An emission line galaxy is an AGN if   
\begin{equation}
\log{{[OIII]\lambda5007}\over{H_\beta}}>
1.3+0.61/\Big(\log{{[NII]\lambda6583}\over{H_\alpha}}-0.05\Big).
\label{agn_eq}
\end{equation}
We selected all SDSS BCD candidates chosen according to \citet{gil03} criteria
(items 1, 2 and 3 in Table~\ref{criteria_bcd}),
and having the spectral line information required to
apply the test 
(the lines $[OIII]\lambda5007$, 
$[NII]\lambda6583$, $H_\alpha$, and $H_\beta$).
Figure~\ref{agn} shows the scatter plot
of the two indexes involved in equation~(\ref{agn_eq}). 
Points above the solid line are AGNs
according to the criterion in equation~(\ref{agn_eq}). 
There are very few AGNs and, what is even 
more important, most of them stay to the
right of the vertical dashed line, i.e., 
\begin{equation}
{\rm N2}=\log{{[NII]\lambda6583}\over{H_\alpha}} > -0.95, 
\end{equation}
which corresponds to the constraint on the N2 index imposed
when the O/H abundance estimate is based on the N2 index, and the O/H 
is constrained as we do (item 5 in Table~\ref{criteria_bcd}). Our BCD candidates are 
to the left of this line and therefore the low O/H abundance constraint 
automatically removes most of the AGN contamination,
which explains our approach. In order to  
quantify the residual contamination, let us mention that 
only 0.4\% of the low O/H abundance BCD candidates in Fig.~\ref{agn} 
are also AGN candidates.

We select isolated BCD candidates (criterion~7 in Table~\ref{criteria_bcd}) 
because of consistency with the
criteria used to search for \bcd\ host galaxy candidates. 
As in the case of \lsb s, we  
stipulate that
the galaxies have no bright 
companion  within $10~R_{50}$, 
where $R_{50}$ is 
the radius 
including 50\% of the Petrosian flux. The 
companions are not bright enough to perturb the galaxy if
they are three magnitudes fainter than the galaxy. 

Applying the criteria in Table~\ref{criteria_bcd} to the SDSS/DR6, 
one gets 1609 \bcd\ galaxies. The dotted lines in Fig.~\ref{myBCD} represent
the distribution of physical properties of this set of candidates.
Means, standard deviations, and modes are also included in 
Table~\ref{properties}.
Four randomly chosen \bcd\ candidates are shown in Fig.~\ref{image_sample},
bottom.
The main redshift of the \bcd\ candidates is 0.032, with a
standard deviation of 0.019. 

After carrying out the selection described above, we figured
out
that most of the \bcd\ sample is also part of 
the \lsb\ sample: 1198 galaxies are shared by the two
sets. They represent 74.5\% of the \bcd s and 5.6\%
of the \lsb s. The typical properties 
of the sets are very different (Table~\ref{properties} and
Fig.~\ref{myBCD}), however  
there is  overlapping between the two populations.
Roughly speaking, the \bcd\ sample represents the 
fraction \lsb\ galaxies having the largest 
$H_\alpha$ equivalent width (see  Fig.~\ref{myBCD}, bottom right,
where the dotted line and the solid line agree for 
$H_\alpha$~EW~$\ga 100$~\AA ).
Since the $H_\alpha$ luminosity is a proxy for 
star formation \citep[e.g.][]{ken98}, 
the \bcd\ sample seems to be the \lsb s
having the largest specific Star Formation Rate (SFR),
i.e., the largest SFR per unit of luminosity.
The overlapping is consistent with the two galaxy sets being part of 
a single continuous sequence, the most active \lsb\ 
galaxies still being identified as 
\bcd\ galaxies.

\section{Luminosity functions}\label{lf}

The LF, $\Phi(M)$, is defined as the number of
galaxies with absolute magnitude $M$ per unit volume and 
unit magnitude. We will use it 
to quantify the number
of \lsb\ candidates per \bcd\ galaxy existing in the
nearby universe.  
LFs are computed using a maximum likelyhood 
procedure similar to the methods 
described in the literature \citep[e.g.,][]{efs88,lin96,tak00,bla01}. 
We developed and tested our own code for 
training purposes, and its characteristics and 
similitude with existing methods are described in 
App.~\ref{appa}. 
Cubic splines are used to parameterize the LF shape,
and the best fit is retrieved maximizing the 
likelyhood in equation~(\ref{loglike}), a task realized
with the usual Powell algorithm \citep[e.g.][]{pre88}. 
Errors bars are assigned by bootstrapping.

\begin{figure*}
\includegraphics[width=0.7\textwidth,angle=90]{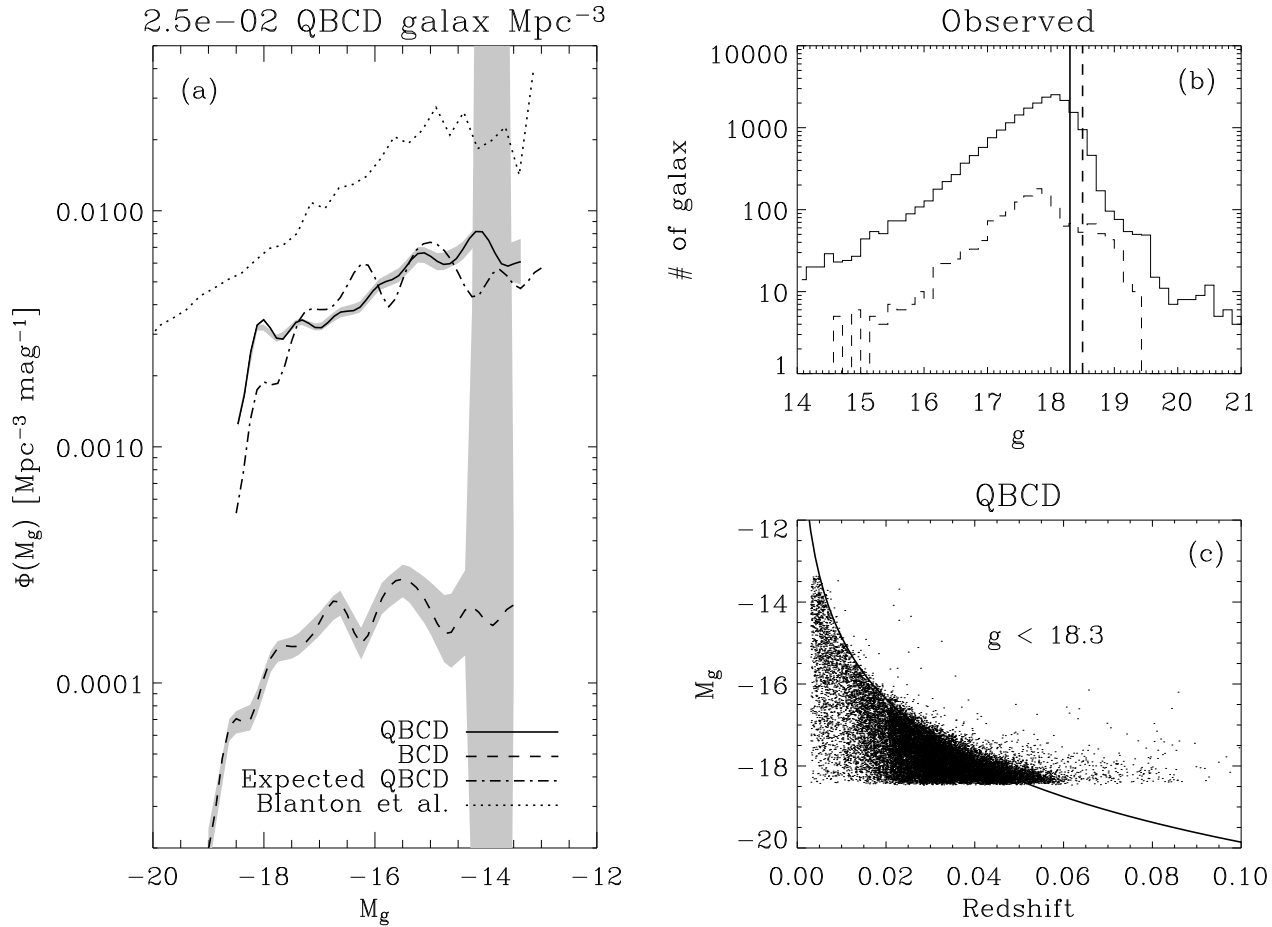}
\caption{
(a) $g$ color LFs of the \lsb\ galaxy sample 
(the solid line), and the BCD galaxy sample (the dashed
line). The shaded area shows our statistical
error estimate. The dotted-dashed line represents
the LF to be expected when the starburst of the BCD 
galaxies fades away revealing the underlying host galaxies. 
The dotted line corresponds to the
LF of low luminosity galaxies
by \citet{bla05}, and it is shown here for
reference.
(b) Observed apparent magnitudes $g$,
with the vertical lines corresponding to the magnitude
limits used to derive the LFs.
As the inset in (a) indicates, the solid line and the
dashed line represent the \lsb\ sample and
the BCD sample, respectively.
(c) Scatter plot of the absolute magnitude of the \lsb\ 
galaxies vs the redshift. The solid line shows the 
boundary to be expected if the sample were limited
in magnitude with $g < 18.3$. 
}\label{lfs_fig}
\end{figure*}

When the procedure is applied to the \lsb\ galaxies 
described in \S~\ref{lsbstuff}, one finds the 
LF represented in Fig.~\ref{lfs_fig}a (the solid line).
This LF is similar to that obtained with the $1/V_{\rm max}$ 
method working
on the same data sets (see App.~\ref{appa}). 
The formal errors deduced from bootstrapping are very small
(the shaded area around the solid line in Fig.~\ref{lfs_fig}a).
We use 50 bootstrap re-samples, but the
error estimate is not very sensitive to this parameter.
The LF is mostly sensitive to the apparent 
magnitude limit of the \lsb\ galaxy set.
We are using a single apparent magnitude limit
for the full dataset, and changing this limit modifies
the overall normalization --
given a number of observed galaxies, the deeper the 
magnitude limit the lower the inferred number 
density of galaxies. We take for the limit, 
\begin{equation}
g< 18.3. 
\label{mlimit}
\end{equation}
This selection is consistent with the magnitude limit
of the main SDSS galaxy sample 
\citep[$r < 17.8$,][]{ade08}, keeping in mind
that the \lsb\ galaxies are somewhat red with $g-r\simeq 0.5$
(see Table~\ref{properties}
and Fig.~\ref{myBCD}). 
However, we obtain the limit in 
equation~(\ref{mlimit}) from the scatter plot of 
the \lsb\ galaxy absolute magnitude vs the redshift shown 
in Fig.~\ref{lfs_fig}c. 
By trial and error, we modify the curve representing
the boundary to be expected if the whole 
sample were limited with a single apparent magnitude. The 
best match is shown as the solid line in Fig.~\ref{lfs_fig}c
and, except for the range of large luminosities, the data set fits 
in well the expected behavior. Galaxies whose apparent magnitudes 
exceed the threshold in equation~(\ref{mlimit}) are not used to compute the LF --
a histogram of the observed apparent magnitudes 
 is shown in Fig.~\ref{lfs_fig}b (the solid line). 
The normalization of the LF depends on the fraction of
sky covered by survey, which 
we take to be the coverage of the SDSS/DR6 spectroscopic 
catalog (7425~deg$^2$; \citealt{ade08}).

Figure~\ref{lfs_fig}a also shows the LF corresponding to the
BCD galaxies selected in \S~\ref{select_bcd} (the dashed line).
The number density of BCD candidates ($n_{0~\rm BCD}\simeq 9.2\,10^{-4}$~Mpc$^{-3}$)
is much smaller than the
number density of \lsb\ galaxies ($n_{0~\rm QBCD}\simeq 2.5\,10^{-2}$~Mpc$^{-3}$),
giving,
\begin{equation}
{{n_{0~\rm QBCD}}\over{n_{0~\rm BCD}}}\simeq 27.
	\label{ration_1_o}
\end{equation}
In this case the limit magnitude of the sample has been set to $g< 18.5$,
which we determine with the same procedure used for the \lsb\ galaxies.
The need for a different  apparent magnitude limit can be inferred
from the histograms in Fig.~\ref{lfs_fig}b. Note how the
histogram for BCD galaxies (the dashed line) does not drop 
off for large magnitudes as abruptly as the histogram for the
\lsb\ galaxies (the solid line). 

\begin{figure*}
\includegraphics[width=0.7\textwidth,angle=90]{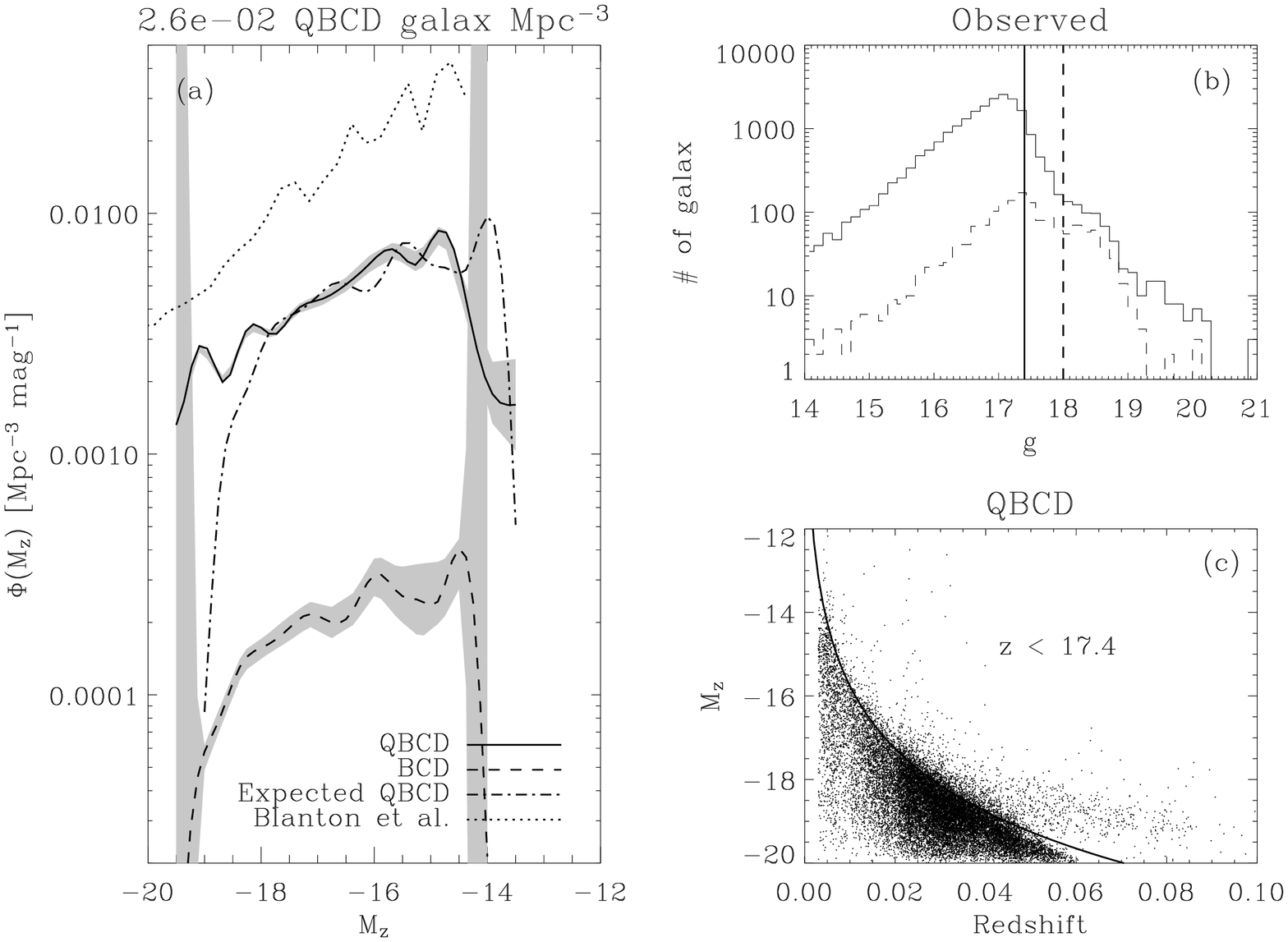}
\caption{
This figure is identical to Fig.~\ref{lfs_fig} except that
the LFs, the histograms, and the scatter plot refer to $z$, 
i.e., the redder color among the SDSS bandpasses.
See Fig.~\ref{lfs_fig} caption for details.
}
\label{lfs_z_fig}
\end{figure*}

The calculus of LFs has been repeated using $z$ magnitudes, i.e.,
the redder SDSS color filter at $9097~$\AA. This bandpass is less
sensitive to (blue) starbursts, enhancing the contribution
of old stellar populations.
The results are represented in Fig.~\ref{lfs_z_fig}.
They are similar to those obtained from $g$, except that
the LFs are shifted by one magnitude. The ratio between
the number of \lsb\ galaxies and BCD galaxies remains as in 
equation~(\ref{ration_1_o}) 
-- 24 in this particular case.


Figures~\ref{lfs_fig}a and \ref{lfs_z_fig}a include the 
LFs for extremely
low luminosity galaxies worked out by \citet{bla05} 
(the dotted lines). We used them as a reference, since
they include all low redshift galaxies in the SDSS spectroscopic 
catalog and, therefore, they provide LFs for the local universe 
affected by the same kind of bias as our galaxy selection. 
Using these LFs as reference, we find the  \lsb\ galaxies  
to be rather numerous. One out of each three  
dwarf galaxies is a \lsb\ candidates
(cf. the solid lines and the dotted lines in 
Figs.~\ref{lfs_fig}a and \ref{lfs_z_fig}a).
Similarly, one out of each ninety  
dwarf galaxies is a \bcd\ candidate\footnote{In a reply to J.~Young,  
	\citet{thu91} estimates the same ratio using 
	very different arguments.}
(cf. the  dashed lines and the dotted lines in 
Figs~\ref{lfs_fig}a and 
\ref{lfs_z_fig}a).

Our LFs are not corrected for the incompleteness of the SDSS catalog at
low surface brightness. According to the detailled
simulation carried out by \citet[][Fig.~3]{bla05},
the incompleteness is not significant for galaxies
with $\langle \mu_g\rangle < 23.5$.
Most of our \lsb\ candidates are within this bound
(Fig.~\ref{myBCD}, bottom left), and it does not affect
the \bcd\ selection at all (Table~\ref{criteria_bcd}).
The \lsb\ LF would be affected at its faint end,
but even at $M_g\sim -14$ the expected 
correction is not larger than a 
factor of two \citep[][Fig.~6]{bla05}.
This uncertainty does not modify the conclusions 
of the paper, which are mostly
qualitative. However,
the number of \lsb\ galaxies worked out in the
paper may be underestimated
by a factor of 
order one.

\subsection{From LF of \bcd s to LF of \bcd\ hosts}\label{bcd2lsb}

The question arises as to whether the LFs for \lsb\ galaxies and 
BCD candidates are consistent. They seem to be, according to the
following heuristic procedure to estimate 
the LF for the hosts of the \bcd s.  
Given a LF of BCDs, $\Phi_{\rm BCD}$, one can work out
the LF to be expected once the 
starburst dies out, $\Phi_{\rm QBCD}$. 
By switching the star formation
on and off,
BCD galaxies turn into \lsb\ galaxies, and vice versa. 
These two 
evolutionary phases have
different timescales; 
 $\tau_{\rm BCD}$ for the BCD phase, and $\tau_{\rm QBCD}$ for 
the \lsb\ phase. In a stationary state the number of
galaxies turning from BCD to \lsb\ must be
balanced by the galaxies going from \lsb\ to BCD,
\begin{equation}
\Phi_{\rm QBCD}(M){{\Delta M}\over{\tau_{\rm QBCD}}}
\simeq \Phi_{\rm BCD}(M'){{\Delta M'}\over{\tau_{\rm BCD}}},
\label{stationary}
\end{equation}
where BCD galaxies with magnitude in the
interval $M'\pm\Delta M'/2$ become  
\lsb\ galaxies with magnitudes $M\pm\Delta M/2$.  
Combining equation~(\ref{stationary}) with
the empirical relation between the magnitudes of the host 
galaxy and
the BCD (equation~[\ref{law}]), one has the recipe 
to infer the LF of the host galaxies to be expected from the 
LF of the BCD galaxies,  
\begin{equation}
\Phi_{\rm QBCD}(M)
\simeq {{\tau_{\rm QBCD}}\over{\tau_{\rm BCD}}}\Phi_{\rm BCD}(M-0.5).
	\label{phi_to_phi}
\end{equation}
The scaling factor $\tau_{\rm QBCD}/\tau_{\rm BCD}$ is just the ratio between 
the number densities of BCDs, $n_{0~\rm BCD}$,
and \lsb s, $n_{0~\rm QBCD}$,
\begin{equation}
{{\tau_{\rm QBCD}}\over{\tau_{\rm BCD}}}\simeq 
{{n_{0~\rm QBCD}}\over{n_{0~\rm BCD}}}, 
\label{timescales}
\end{equation}
since $\Phi_{\rm QBCD}$ and  $\Phi_{\rm BCD}$ are normalized to 
$n_{0~\rm QBCD}$ and $n_{0~\rm BCD}$, respectively.

Using equations~(\ref{phi_to_phi}) and (\ref{timescales}), 
we evolve the LF of BCD galaxies in Figs.~\ref{lfs_fig}a and 
\ref{lfs_z_fig}a to obtain the LFs of their host galaxies. 
They are included in the same figures as the dotted-dashed lines. 
The scaling~(\ref{timescales}) has been taken as the
ratio between the number densities of \lsb\ galaxies and
BCD galaxies (i.e., the ratio in equation~[\ref{ration_1_o}] for 
the $g$ filter, and the corresponding
figure for the $z$ filter).
The similarities between the evolved LF (the dotted-dashed lines) 
and the LF computed
from the \lsb\ data sets (the solid lines) are quite striking. 
We interpret this  agreement as an indication of 
self-consistency between the two
sets of galaxies chosen in \S~\ref{data_set}. BCD galaxies and 
\lsb\ galaxies can be different phases in the life 
of a dwarf galaxy, with the lifetime in the 
BCD phase some thirty times  shorter than the lifetime in the \lsb\ phase
(equations~[\ref{ration_1_o}] and [\ref{timescales}]).

The \bcd\ starbursts are
very young since they still
conserve massive stars (see \S~\ref{introduction}). If 
$\tau_{\rm BCD}\simeq 10~$My then 
equations~(\ref{timescales}) and (\ref{ration_1_o}) imply
$\tau_{\rm QBCD}\simeq 0.27~$Gy. Consequently, each \lsb\ galaxy
may undergo as many as 30-40 star formation episodes
during the timespan  where
stars can be formed in  dwarf galaxies 
\citep[$\ga$10 Gy; e.g.,][]{kun00}.
This issue is discused in \S~\ref{discusion}.

\section{Properties of {\lsb} galaxies and {\bcd} galaxies}\label{malmquist}
The histograms of galaxy properties discussed
in \S~\ref{data_set}, and shown in Fig.~\ref{myBCD},
describe the properties of the {\em observed} galaxies. 
These histograms are strongly biased since they overweight 
the properties the 
most luminous 
galaxies in the samples. In order to correct for
the Malmquist bias, so that histograms are weighted according to 
the {\em true} number density of galaxies, we have used the ratio 
between the LF  derived in \S~\ref{lf}, $\Phi(M)$, 
and the observed histogram of absolute magnitudes,
\begin{equation}
h(M)={1\over{\Delta M}}\sum_i\,\Pi\big({{M_i-M}\over{\Delta M}}\big).
\label{pi_eq}
\end{equation}
As usual, the symbol $\Pi$ stands for the
rectangle function,
\begin{equation}
\Pi(x)=\cases{
1&$|x|<1/2$,\cr
0&elsewhere.
}
\end{equation} 
The index $i$ in equation~(\ref{pi_eq}) includes all observed 
galaxies, whereas $M$ and $\Delta M$ determine
the centers and the widths of the histogram bins
\footnote{Note that the histogram defined in equation~(\ref{pi_eq}) differs
from that used in Fig.~\ref{myBCD} because of the factor $\Delta M^{-1}$.
This trivial re-scaling is used here for convenience, allowing
the corrected histograms to be normalized to the 
number density of galaxies.
}.
We have to assume that each observed galaxy is actually 
a proxy for $b_i$ galaxies, $b_i-1$ of which are not 
included in our observation because they are too 
faint to exceed the apparent magnitude threshold. 
The magnitude of the galaxy causes the bias and, therefore,
we assume that the bias function only depends on the
absolute magnitude of the galaxy, 
\begin{equation}
b_i=b(M_i).
	\label{bias_funct}
\end{equation}
This bias is precisely the reason why  
$h(M)$ is not the LF, therefore, for $\Delta M$ small enough,
\begin{equation}
\Phi(M)\simeq {1\over{\Delta M}}\sum_i\,b_i\,\Pi\big({{M_i-M}\over{\Delta M}}\big).
	\label{phi_h}
\end{equation}
Using equation~(\ref{bias_funct}),
and assuming that
$b(M)$ is a slowly varying function of $M$,
\begin{equation}
b(M_i)\,\Pi\big({{M_i-M}\over{\Delta M}}\big)\simeq
b(M)\,\Pi\big({{M_i-M}\over{\Delta M}}\big),
\end{equation}
and so 
one finds from equation~(\ref{phi_h})
the expression for the bias function,
\begin{equation}
b_j=b(M_j)=\Phi(M_j)/h(M_j).
\label{b_i_est}
\end{equation}
Let us denote as $k(p)$ the histogram of any parameter $p$
constructed using the observed galaxies,
\begin{equation}
k(p)={1\over{\Delta p}}\sum_i\,\Pi\big({{p_i-p}\over{\Delta p}}\big),
\end{equation}
with $\Delta p$ the bin-size and $p_i$ 
the value corresponding to the $i$-th galaxy.
Then the bias-corrected histogram would be,
\begin{equation}
k^*(p)={1\over{\Delta p}}\sum_i\,b_i\,\Pi\big({{p_i-p}\over{\Delta p}}\big).
\label{correct_histo}
\end{equation}
This definition guarantees that the corrected
histogram of observed absolute magnitudes is the luminosity function 
(cf. equations~[\ref{phi_h}] and [\ref{correct_histo}] with $p=M$)
and, therefore, it is not difficult to show
that all corrected histograms
have the same normalization as the LF, i.e.,
\begin{equation}
\int_{-\infty}^{\infty}k^*(p)\,dp=n_0,
\end{equation}
$n_0$ 
being
the number density of galaxies (see App.~\ref{appa}).

\begin{figure*}
\includegraphics[width=0.7\textwidth,angle=90]{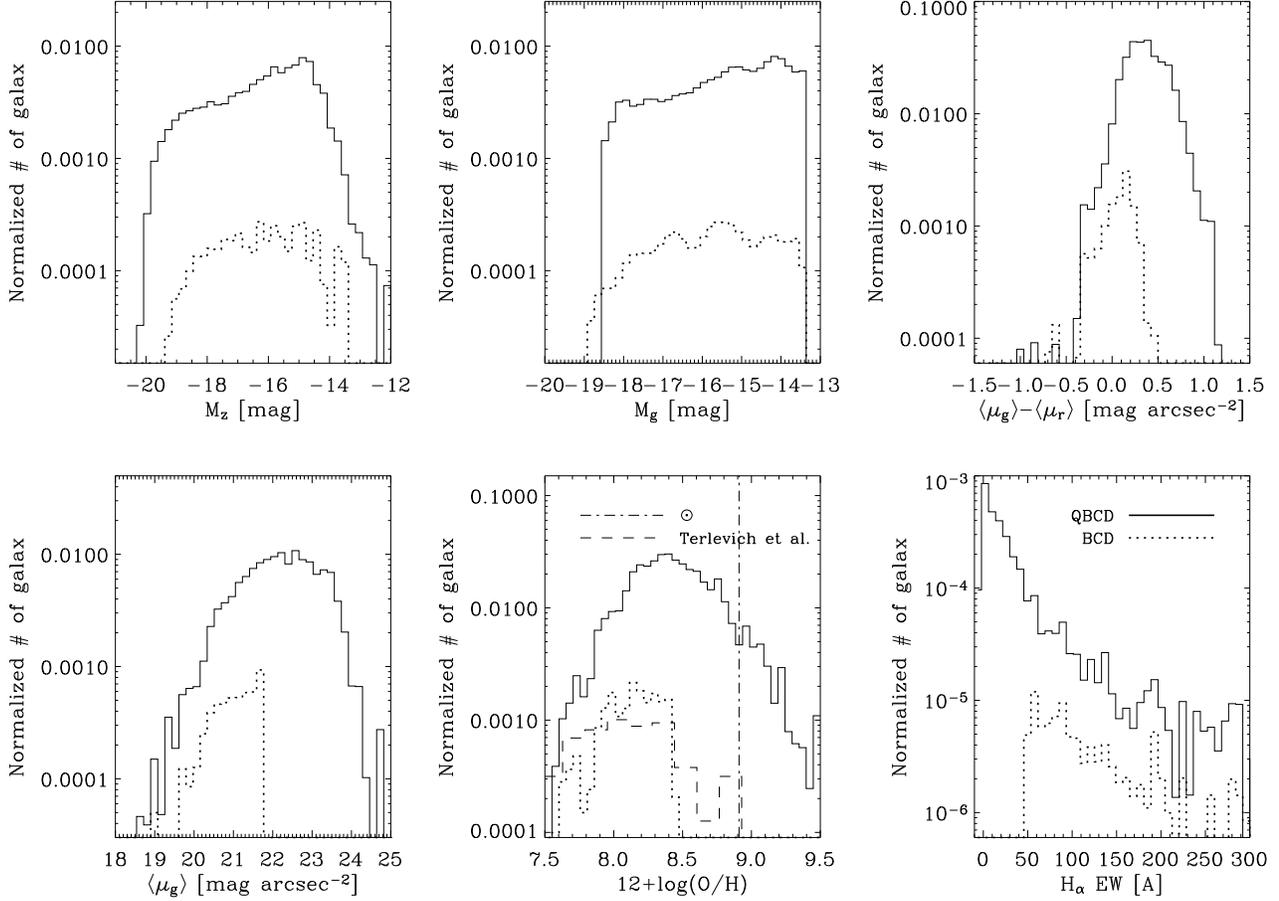}
\caption{
Histograms of magnitudes and colors for \lsb s (the solid lines) and \bcd s
(the dotted lines) once the Malmquist bias has been compensated for.  
The unrestored versions of these histograms are shown in Fig.~\ref{myBCD}.
The labels in the ordinate axes describe the parameter that is represented.
The histograms are normalized to the number density of galaxies.
The dashed line in the plot of metallicities ($12+\log [{\rm O/H}]$)
reproduces Fig.~3 in \citet[][]{kun00},
which corresponds to the abundances of 425 H~{\sc ii} galaxies
cataloged by \citet{ter91}. Refer to Fig.~\ref{myBCD} for further details.}
\label{myBCD_restored}
\end{figure*}

\begin{figure}
\includegraphics[width=0.5\textwidth]{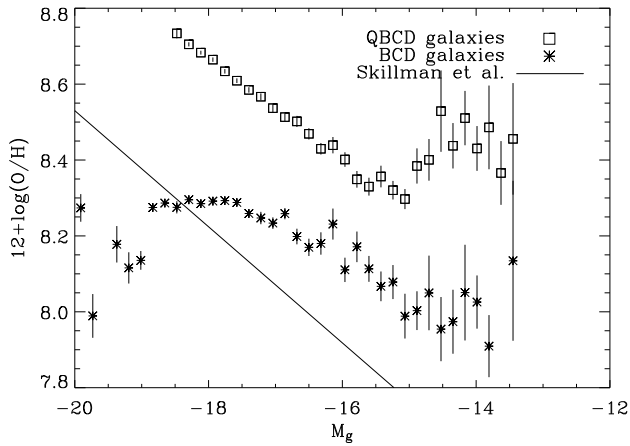}
\caption{
Metallicity vs absolute $M_g$ magnitude.
The symbols stand for the mean values considering all the galaxies 
with similar $M_g$, whereas the error bars represent 
the standard deviation of the mean values.
Both \lsb\ galaxies (the square symbols) and BCD galaxies (the 
asterisks) follow a trend so that the fainter the galaxy the more metal poor.
The solid line is shown for reference and it corresponds to
the law for nearby dwarf irregular galaxies found by \citet{ski89}. 
}\label{abu_vs_mg}
\end{figure}

\begin{figure}
\includegraphics[width=0.5\textwidth]{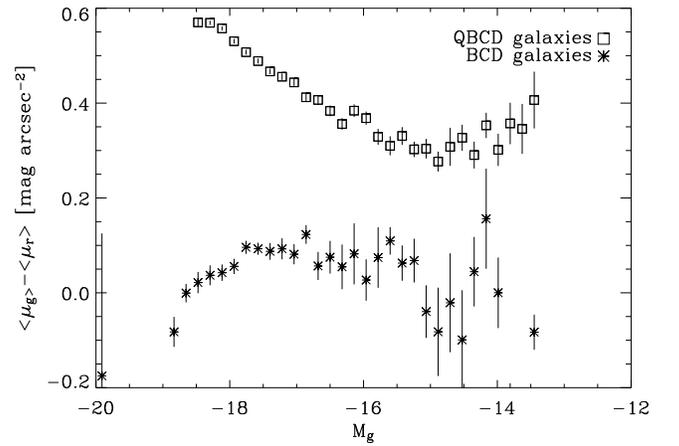}
\caption{
$\langle\mu_g\rangle -\langle\mu_r\rangle$ color vs absolute $M_g$ magnitude.
The symbols stand for the mean values considering all the galaxies 
with similar $M_g$, whereas the error bars represent  the 
standard deviation to be expected for these mean values.
The \lsb\ galaxies  are bluer as they become
fainter (the square symbols). The same trend is not so
obvious for the BCD galaxies
(the asterisks).
}\label{color_vs_mg}
\end{figure}

\begin{figure}
\includegraphics[width=0.5\textwidth]{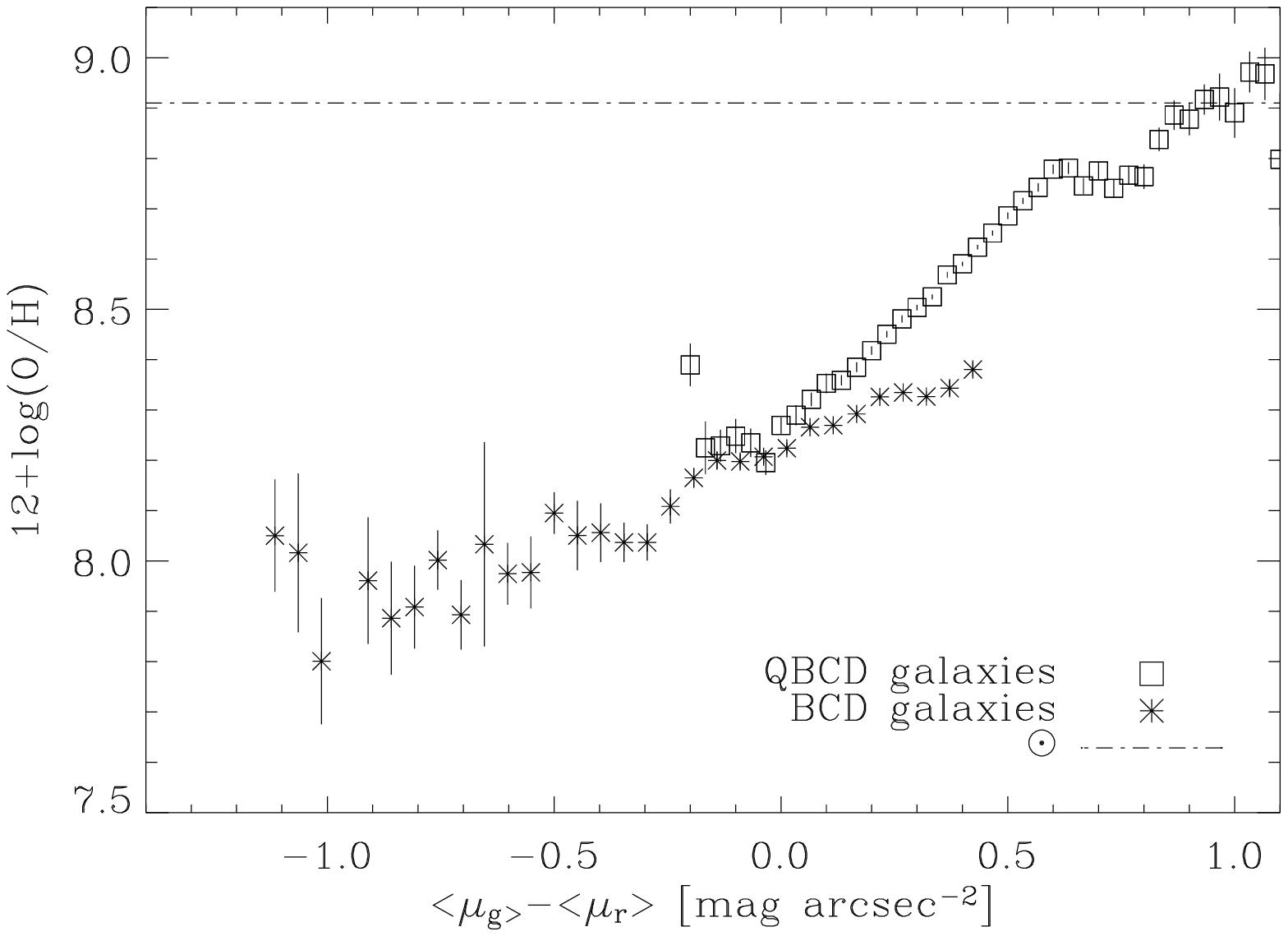}
\caption{
Metallicity vs color. There is a clear trend, which
becomes  steeper and 
sharper in the case of the \lsb\ galaxies (the square symbols).
The solar metallicity has been marked for reference. 
Each symbol represents the average metallicity ($12+\log[{\rm O/H}]$)
considering all the galaxies with the same color. The error
bars correspond to the standard deviation to be expected for these averages.
}\label{abu_vs_color}
\end{figure}

We have applied equations~(\ref{b_i_est})
and (\ref{correct_histo}) to restore the histograms
in Fig.~\ref{myBCD}. The result is shown in Fig.~\ref{myBCD_restored}.
The means, standard deviations and modes of the new histograms are also
included in Table~\ref{properties}. As expected, the mean 
absolute magnitudes of the restored histograms are much fainter 
than the observed ones. 
However, in addition to this effect, the 
corrected histograms hint at intrinsic metallicities significantly lower 
than the observed ones both for BCD and \lsb\ galaxies.
The corrected \lsb\ galaxy colors are bluer than
the colors of the observed set. 
These  changes can be pin
down to the correlations between luminosity, color, 
and metallicity existing in the original data set.
Figure~\ref{abu_vs_mg} shows how the \lsb\ galaxies
tend to be less metallic as they become fainter,
and a similar, but less marked trend, is also present
in BCD galaxies. Such a relationship between metallicity and
mass in dwarf galaxies
(and so, between metallicity and luminosity) 
was given by \citet{pag81}, and later on by
many others \citep[see][]{kun00}.
Figure~\ref{abu_vs_mg} includes the linear relationship 
found by \citet{ski89} for nearby dwarf irregular galaxies.
Curiously enough, the slope is almost the same 
as we find for
\lsb\ galaxies.
There is also a relationship between color and luminosity, so that
fainter \lsb\ galaxies
seem to be bluer; see Fig.~\ref{color_vs_mg}.
Figures~\ref{abu_vs_mg} and \ref{color_vs_mg} actually reveal an 
underlying relationship between color and metallicity. It has
been brough up in Fig.~\ref{abu_vs_color}, showing how 
the redder the galaxy the larger the oxygen abundance.
The relationship is more clear in the case of the
\lsb\ galaxies (the square symbols), but it is also
present among the redder BCD galaxies (the asterisks
for $\langle\mu_g\rangle-\langle\mu_r\rangle > -0.4$).
The relationships for \lsb s and  BCDs meet at 
$\langle\mu_g\rangle-\langle\mu_r\rangle\simeq -0.1$,
although they have different slopes. 

The number of known BCD galaxies with metallicities less than 
1/20 the solar value is rather small 
\citep[a dozen or so, according to ][]{kun00}. Surprisingly enough,
the list of 1609 BCD candidates we work out
contains only three candidates below this metallicity. 
We interpret this result as a
support of the metallicity estimate 
carried out in the paper,
but it also reinforces the existence of a
minimum H~{\sc ii} region based metallicity 
\citep[see][and references therein]{kun00}.
%

\begin{figure}
\includegraphics[width=0.5\textwidth]{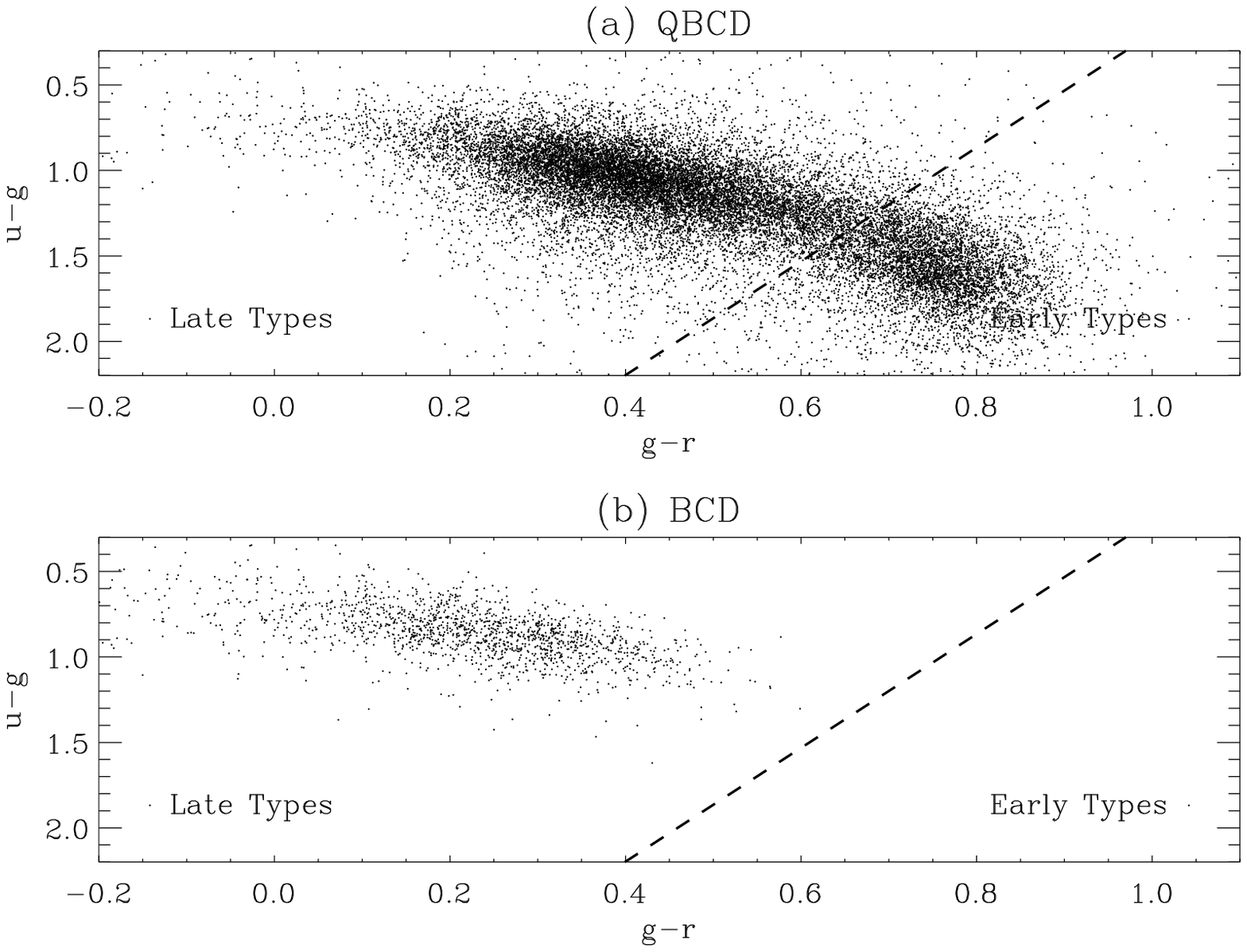}
\caption{
(a)~$u-g$ vs $g-r$ scatter plot for the \lsb\ candidates.
The colors in the sequence span from those of early-type
galaxies, to those of late-type galaxies. There seem to be
two clusters in the distribution of points, 
which we have artificially separated by the dashed line.
(b)~Same as (a) but showing the \bcd\ candidates.  
}
\label{color_vs_color}
\end{figure}

The range of selected \lsb\ colors is rather broad and, therefore, 
our list of \lsb\ candidates seems to include all the range
from early-type galaxies to late-type
galaxies. 
This fact can be appreciated in Fig.~\ref{color_vs_color}a,
which contains a color-color scatter plot.
Our galaxies follow the sequence used in galaxy classification,
with the reddest extreme corresponding to elliptical galaxies (early types),
and the bluest extreme to irregular galaxies (late types) 
-- 
compare Fig.~\ref{color_vs_color}a with, e.g., Fig.~2 in \citet{ber00}.
One can distinguish two clusters or concentrations in 
Fig.~\ref{color_vs_color}a (the dashed line 
arbitrarily separates the two sub-sets). 
The existence of these two clusters or  
classes was to be  
expected, since they correspond to the 
bimodal distribution of colors found in the large samples of 
galaxies \citep[e.g.,][]{bal04,bla05}. 
Hints of these two modes are also found as two small peaks
in the histogram of surface brightness color shown in Fig.~\ref{myBCD}
(top right, the solid line, with the two peaks at 
$\langle\mu_g\rangle-\langle\mu_r\rangle\simeq$ 0.4 and 0.7). 
Note how the \bcd\ candidates occupy the bluest extreme of the color-color 
plot; cf. Figs.~\ref{color_vs_color}a  and \ref{color_vs_color}b. 

Most of the \lsb\ candidates show $H_\alpha$ in 
emission, meaning that even if the star-formation
is reduced with respect to the \bcd\ candidates,
it is not absent. Only some 3200 \lsb\ galaxies
out of the 21493 candidates show $H_\alpha$
in absorption. They have (negative)
equivalent widths of a few \AA\ and, therefore, 
these objects are  included in the bin of zero 
equivalent width in Fig.~\ref{myBCD}, bottom right.
Using the $H_\alpha$ flux as a proxy for
star formation \citep[][]{ken98}, 
our \lsb\ candidates turn out
to be one order of magnitude less active 
in forming stars than the \bcd s.
This estimate is worked out in \S~\ref{consumption},
equations~(\ref{comp6}) and (\ref{comp5}).

\section{Discusion}\label{discusion}

So far we have described our search avoiding 
interpreting the results. This section, however,
is fully devoted to interpretation 
and so, admittedly speculative.
We will discuss
how the results in previous sections are consistent
with the BCD candidates and the \lsb\ candidates  
being two different phases in the same sequence of 
galactic evolution.
Using as a working hypothesis that \bcd s change 
into \lsb s, and vice~versa, we examine obvious
flaws and constraints.
Showing the consistency of an 
hypothesis does not prove it to be correct -- 
other alternatives cannot be discarded. 
It just shows the working hypothesis to be viable.
Three main results point out in this direction.
First, we choose the \lsb s so that their colors, concentration
indexes and luminosities are equal to those of the
galaxies underlying the \bcd s.
Second, the LF of \lsb s is conformable to the LF of
\bcd s (\S~\ref{lf}).  Third, the \bcd\ galaxies turn out to be
\lsb\ galaxies with the largest (specific) star formation
rates (\S~\ref{select_bcd}).  

Matching  LFs forces the \lsb\ phase to last some 30 times 
longer than the BCD phase (\S~\ref{bcd2lsb}). Since the BCD
starbursts are so short-lived, there should be many 
episodes of BCD phase during a galaxy lifetime. Assuming a 
single 10~My long starburst per BCD phase, there should
be one of such BCD phases each 0.3 Gy. 
Longer BCD phases are in principle 
possible by concatenation of 
starbursts, but population synthesis modeling of BCD
spectra does not favor extended periods of star-formation 
\citep[e.g.,][]{mas99}. 
%
If \bcd s are transformed into \lsb s and vice versa,
each \lsb\ has experienced several \bcd\ phases  
because the other alternative does not seem to be consistent
with the results of the search.
Let us examine this alternative 
possibility, namely, that the BCD phase represents the only episode 
of intense  star formation during the galaxy lifetime. 
Statistically, all \lsb\ galaxies that we
detect today have sufferred a BCD episode during the
last 270~My (equations~[\ref{ration_1_o}] and [\ref{timescales}], 
with $\tau_{\bcd}=$10~My). If this is the only 
such
episode per galaxy, 
there should be no BCDs among galaxies observed at lookback
times larger than 270~My, or at redshifts  $> 0.02$ (= $270~{\rm My}/H_0$). 
However, a large fraction of the sample of BCDs selected 
in \S~\ref{select_bcd} has redshifts larger than this limit.
We are forced to conclude that the same host 
galaxy undergoes several BCD phases.

Although the recursive BCD phase scenario explains 
a number of important observables, 
other properties of \lsb s and \bcd s 
do not seem fit in the picture so well.
The purpose of the rest of the section is to discuss 
(and hopefully  clarify) some of the most obvious 
difficulties posed by the scenario.

\subsection{H~{\sc i} consumption timescale}\label{consumption}

One may think that the repeated starbursting
of a dwarf  galaxy
quickly exhausts the gas reservoir,
thereby
making the whole picture inconsistent.
However, the low surface brightness galaxies 
like the \lsb s are gas-rich 
\citep{sta92,kun00}, with fuel to power the
star formation during several Hubble times.
This claim 
can be supported by working out
the H~{\sc i} consumption timescale, 
$\tau_{{\rm H\,I}}$, which is defined as the 
time required to transform into stars the  
neutral hydrogen of the galaxy, $M_{{\rm H\,I}}$, i.e.,
\begin{equation}
\tau_{{\rm H\,I}}=M_{{\rm H\,I}}/{\rm SFR},
\label{def_tau}
\end{equation}
with  SFR the  star formation rate.
On the one hand,
dwarf low surface brightness galaxies have about one solar mass of 
H~{\sc i} per solar luminosity \citep{sta92},
explicitly,
\begin{equation}
\log(M_{{\rm H\,I}}/M_\odot)=-0.4\,(M_g-M_{g\,\odot}),
\label{comp1}
\end{equation}
$M_\odot$ 
being one solar mass, and $M_{g\,\odot}$ the solar
absolute luminosity in the $g$ filter.
On the other hand, the SFR scales with the $H_\alpha$
luminosity \citep[e.g.,][]{ken98}, i.e., with the product of the $H_\alpha$
equivalent width times the galaxy luminosity. Since the observed
BCD $H_\alpha$ equivalent width is uncorrelated with the galaxy 
luminosity, the BCD SFR scales with the
luminosity, that is to say,
\begin{equation}
\log({\rm SFR_{BCD}/SFR'_{BCD}})=-0.4(M_g-M_g'),
\label{comp2}
\end{equation}
where ${\rm SFR'_{BCD}}$ stands for the SFR of a BCD 
galaxy of magnitude $M_g'$. In our two-phase  scenario, the galaxy
has short periods of star formation lasting $\tau_{\rm BCD}$,
interleaved  with long periods of quiescence lasting $\tau_{\rm QBCD}$
(\S~\ref{bcd2lsb}).
Consequently, the effective time-average
SFR is, 

\begin{equation}
{\rm SFR}\simeq
{
{\tau_{\rm BCD}\,{\rm SFR_{BCD}}+
\tau_{\rm QBCD}\,{\rm SFR_{QBCD}}
}
\over
{
\tau_{\rm BCD}+\tau_{\rm QBCD}}}
\label{comp4}
\end{equation}
\begin{displaymath}
\simeq 
{\rm SFR_{BCD}}
\Large(
{{\tau_{\rm BCD}
\over
{\tau_{\rm QBCD}}}}
+\epsilon
\large),
\end{displaymath}
with
\begin{equation}
\epsilon =
{
{\rm SFR_{QBCD}}
\over
{\rm SFR_{BCD}}
}.
\label{comp6}
\end{equation}
As judged from the ratio of $H_\alpha$ equivalent widths in
Table~\ref{properties}, and the difference of luminosity
between \bcd s and \lsb s
(equation~[\ref{law}]),
the SFR during the \bcd\ phase is some ten times 
larger than during the \lsb\  phase, i.e.,
\begin{equation}
\epsilon^{-1}\simeq 10.
\label{comp5}
\end{equation}
By combining
equations~(\ref{def_tau}), (\ref{comp1}), (\ref{comp2}) 
and (\ref{comp4}),
one finds the 
consumption timescale to be independent of the galaxy luminosity,
\begin{equation}
\log\tau_{{\rm H\,I}}\simeq -0.4(M_g'-M_{g\,\odot})
-
\log\Big[{{\rm SFR_{BCD}'}\over{M_\odot}}
\big({{\tau_{\rm BCD}}\over{\tau_{\rm QBCD}}}+\epsilon\big)\Big].
\label{comp3}
\end{equation}
Although
BCD star formation bursts are intense for 
a dwarf galaxy,  the SFRs of  BCDs are rather modest, i.e.,
less than  $1 M_\odot {\rm y}^{-1}$, and typically
one order of magnitude smaller \citep[e.g.][]{sag92,mas99}. Taking 
${\rm SFR}'_{\rm BCD} < 1 M_\odot {\rm y}^{-1}$ for the brightest galaxies of
our sample, $M_g'\simeq -19$, and using the ratio of timescales
between the phases as provided by equations~(\ref{ration_1_o}) and (\ref{timescales}),
equations~(\ref{comp5}) and (\ref{comp3}) yield,
\begin{equation}
\tau_{{\rm H\,I}} \ga 
30~{\rm Gy}.
\label{comp9}
\end{equation}
Consequently, the hydrogen existing in \lsb\ galaxies
allows the sequence \bcd -\lsb\ to last for
a few
Hubble times ($\equiv 1/H_0=$14 Gy).

A final comment is in order. The scenario of a 
recursive starbursting  is 
not in conflict with the commonly accepted view of a SFR 
decreasing during the last 10 Gy \citep{mad96,lil96}.
This drop refers to massive galaxies. However,
the kind of dwarf galaxies in our \lsb\ sample,
with stellar masses less than $10^{10}~M_\odot$, 
has maintained a SFR either stationary or
slightly increasing with time 
\citep[see][ Fig.~1]{hea04}.

\subsection{Metallicities}\label{metallicity}

The metallicities of our \lsb\ galaxies are systematically larger
than the metallicities of the \bcd\ galaxies. This result seems to 
be in
conflict with the recursive BCD phase scenario, since 
the BCD starburst starts off with a metallicity lower than
the metallicity of its host galaxy (see Table\ref{properties}). 
Although this disagreement may 
indicate a real flaw in the overall
picture, one can also think of various ways to 
circumvent the apparent inconsistency.  For example,
the BCD episodes may involve fresh low metallicity gas
accreted by the host galaxy during the periods of  
quiescence. Note that the BCD episodes are very conspicuous 
from the point of view of the luminosity, but they involve 
moderate masses, which hardly exceed $10^6\ M_\odot$ 
\citep[e.g.][]{mas99}. A sustained infall 
such as those
postulated in the literature to solve
various problems in galaxy evolution 
($\sim\,1 M_\odot~{\rm pc}^{-2}~{\rm Gy}^{-1}$;
see \citealt{dal07}, and references therein) provides 
$10^6\ M_\odot$ in only 50~My, even for a small 2.5 kpc  
galaxy. Masses of $10^6\ M_\odot$ are also typical 
of the high velocity clouds hounding our galaxy \citep[e.g.,][]{wak07}.
If these gas reservoirs are  common in \lsb s, they
may develop major star formation events when merging with the 
galaxy.

The fresh gas  
glowing
during the 
\bcd\ phase does not necessarily have 
to come from outside. Metal-poor gas may exist in place in the 
galaxy (\S~\ref{consumption}). 
Then the oxygen metallicity assigned here to \lsb\ galaxies 
may not reflect the metallicity of this gas, 
but the metallicity of a few metal-polluted H~{\sc ii} 
regions remaining from the last starburst. 
It is relatively easy to enrich
 with metals the small fraction of galactic gas undergoing 
starbursts. For example, the metallicity 
of a pure gas cloud increases from the level observed in  
\bcd s ($12+\log(O/H)\simeq 8.24$; Table~\ref{properties}) 
to the level in \lsb s ($\simeq 8.61$), if 60\% of the 
original mass is  transformed into 
starts.\footnote{
This estimate assumes the closed box evolution of a 
purely gaseous cloud \citep[e.g.][]{tin80} with
the standard oxygen yield 
\citep[$\sim$0.003; see, e.g.,][]{pil04}.
}
The metal-polluted gas produces emission lines 
during the extended recombination phase of the  H~{\sc ii} 
region \citep[e.g.,][]{bel82}, and it may also give rise to  
a secondary generation of stars. The light from these 
aging  H~{\sc ii} regions renders high metallicity 
measurements.
Eventually, the star-processed gas 
mingles with the metal-poor gas of the galaxy, 
but this mixing leaves
the original metallicity almost 
unchanged\footnote{If 1\% of gas with 
\lsb\ metallicity is mixed up with 99\% of gas 
with \bcd\ metallicity,
then the resulting change of metallicity is 
only $\Delta[12+\log(O/H)]\simeq 0.006$, i.e.,  
insignificant.}. 
We note that this explanation may be in conflict
with the IR excess detected
in the halos of Blue~Compact~Galaxies 
by \citet[][]{ber02}. 
These halos can be identified with the 
\bcd\ host galaxies and, therefore,
with the \lsb s. 
The colors cannot be explained with a normal 
metal-poor stellar population like the Milky Way 
halo, but an excess of low-mass stars with significant 
metallicity is required \citep[see][]{zac06}. 
The nature of the disagreement requires further 
investigation, and it may be due to the fact 
that these Blue Compact Galaxies with  
IR excesses are not dwarfs according to 
the criteria used in our work.

There is yet another possibility to reconcile
the metallicities of the two galaxy types. 
Only those \lsb\ candidates 
with metallicity similar to that of the \bcd\ galaxies
would be part of the 
\bcd\ -- \lsb\ sequence. 
The \lsb s with metallicity like the \bcd s represent 20\% 
of the sample (see Fig.~\ref{myBCD}) and, therefore, the 
timescale of quiescence must be shortened by a factor of
five to comply with equation~(\ref{timescales}). The same
happens with the H~{\sc i} consumption timescale worked out 
in equation~(\ref{comp9}). These \lsb\ galaxies may keep their
low metallicity levels because the metals are 
expelled by galactic winds 
(\citealt{hec02,ten03};
see, however, \citealt{sil01,leg01}),
%
because they are recycled into new stellar generations  within superstar
clusters,  
without significant
mixing with the galactic gas \citep[][]{ten05},
or simply because of the dilution with pristine gas described
above.

\subsection{Starburst triggering}

We select isolated \lsb\ galaxies, therefore,
the starburst triggering mechanism to turn them
into
\bcd s cannot be galaxy-galaxy interactions, 
mergers or  harassment. 
How, then,
are the \bcd\ bursts triggered? 
Several possibilities are available. 
If the fresh gas falls on the \lsb\ galaxy over extended 
periods of time, intense star formation cannot be triggered 
until the gas density exceeds the required 
threshold (the so-called Schmidt law; \citealt{sch59}).
Waiting for enough gas to accumulate could explain  
periods of latency. 
If the gas falls in during short episodes (like the
collision with a large gas cloud), the hitting 
of the gas itself may induce star formation. Then
the \lsb\ timescale would be given by the characteristic
time between cloud collisions. 
If rather than coming from outside,
the fresh gas is  part of the galaxy,  internal galactic
structures like bars may periodically 
excite star formation. In the case of a 
gas-rich galaxy the 
triggering may also be due to perturbations of the 
galactic material by dark matter clumps.
These elusive structures are predicted in lots by
the cold dark matter simulations of galaxy formation 
\citep[e.g.,][]{die07}, and they are expected
to disturb the quiet evolution of galactic
disks \citep[e.g.,][]{kaz07}.

\subsection{\lsb\ characterization criteria}\label{characterization}
The criteria used to characterize 
\lsb s come from the \bcd\ host galaxies
analyzed by \citet{amo07,amo07b}, but we may have 
used different criteria taken from other works. 
The question arises as to whether the selection of 
candidates critically depends on this assumption.

The \bcd\ host galaxy properties are used to
constrain colors and Sersic indexes 
(see \S~\ref{lsbstuff} and Table~\ref{properties}). 
The lower limit luminosity, also taken from the  
\bcd\ host galaxies, turns out to be 
unimportant since the small number low 
luminosity objects prevents any serious 
influence of this criterion on the selection 
(see Fig.~\ref{myBCD}).
The range of visible $g-r$ colors that we use
agrees with the colors that different
studies have assigned to the hosts of \bcd s 
\citep[e.g.,][]{pap96b,gil05}, even in those cases where
a significant infra-red excess has been 
detected  \citep[e.g.,][]{ber02}.  In this sense the  
color criterion does not seem to be 
questionable and, therefore, any 
work would have rendered similar \lsb\ candidates. 
The selection of Sersic indexes 
is more influential. 
Studies like \citet{ber02}
or \citet{gil05} find indexes as large as 
$n=$10--20, whereas we take $n < 3$ 
(Table~\ref{properties}). 
However, there are two good 
reasons to prefer low$-n$ values. First and most 
important, the observed dwarf galaxies present 
low$-n$ values \cite[e.g., Fig.~11 in ][]{cao05,gra03},  
and if we search for galaxies that may be 
Blue-Compact-{\em Dwarf} galaxies during quiescence, 
they must be dwarfs too (see equation~[\ref{law}].)  
Second, the large $n$ values are obtained from 
1D fits to azimuthally averaged luminosity 
profiles, 
and the $n$ thus obtained depends 
critically on the range of 
radii used to characterize the  
galactic outskirts \citep[e.g.,][]{cai03,gil05}.
The 2D fits providing low indexes are fairly more 
robust and, therefore, to be preferred \citep[see][]{amo07}.
%
%
%

\section{Conclusions}\label{conclusions}

The starburst characteristic of
\bcd\ galaxies does not last for long.
Once this episode is over, 
the \bcd\ galaxies should
dim their surface brightness to 
become quiescent \bcd\ galaxies (or \lsb\ galaxies).
Although \lsb\ galaxies are to be expected,  
they have not been identified yet (see \S~\ref{introduction}).
The present work describes an effort to find them among 
the galaxies in the SDSS/DR6 database. The properties of the 
\lsb s have been taken from the sample of \bcd\ host 
galaxies characterized by \citet{amo07,amo07b} (\S~\ref{lsbstuff}). 
We find 21493 \lsb\ candidates, therefore,
\lsb s seem to be  fairly common in the local 
universe. In order to have a proper reference to 
compare with, a complete sample of \bcd\ galaxies was 
selected too. It 
comprises 1609 \bcd\ candidates (\S~\ref{select_bcd}).
Since the two samples were selected from the same 
database using analogous criteria, 
the comparison between the two sets 
is relatively free from the bias that our
selection may have.
 
We compute Luminosity Functions (LFs) for the 
two samples (\S~\ref{lf}).
In addition, we estimate their main properties before and
after correcting for  Malmquist bias. These properties
can be summarized as follows,
\begin{itemize}
\item There are around 30 \lsb\ candidates per \bcd\ candidate.
	We infer this ratio by comparison of the 
	LFs for \lsb s and \bcd s. The two LFs are 
	very similar, except for the global scaling factor,
	and the expected 0.5 magnitude dimming (\S~\ref{bcd2lsb}).
\item  The surface brightness of the \lsb s is 
	typically one
	magnitude fainter than the surface brightness of \bcd s
	(Table~\ref{properties}).  
\item \lsb\ candidates are, on average, 0.4  magnitudes redder
	than the \bcd\ (Table~\ref{properties}; $\langle\mu_g\rangle
	-\langle\mu_r\rangle$).
\item \lsb\ candidates have an H~{\sc ii} region based 
	oxygen metallicity 0.4 dex
	higher than the \bcd\ candidates 
	(Table~\ref{properties}). 
\item The \lsb\  metallicity increases with the luminosity,
	following the well known trend for dwarf galaxies.
\item  The \lsb\  metallicity also increases with the color,
	so that the redder the galaxy the larger the measured
	 metallicity. 
	%
\item 75 \% of the \bcd\ candidates are
	also part of the \lsb\ sample (\S~\ref{select_bcd}). 
	Roughly speaking, the \bcd\ sample represents the 
	fraction \lsb\ galaxies having the largest 
	specific SFR (SFR per unit of luminosity).

\item There are around three
	 dwarf galaxies per \lsb\ candidate
	(\S~\ref{lf}), 
	which renders one \bcd\ galaxy every ninety dwarf 
	galaxies.
\end{itemize}

The overlap 
between \bcd\ galaxies  and \lsb\ galaxies is consistent with 
the two sets forming a single continuous sequence, with the most  
active \lsb\ galaxies being \bcd\ galaxies.
The agreemet between their LF shapes, and the ratio of number their 
densities, support the commonly accepted view that \bcd\ galaxies
undergo short bursts of star formation separated by 
long quiescent epochs. 
However, the fact that the \lsb\ metallicity is
higher than the \bcd\ metallicity poses a problem
to such an
episodic starbursting scenario, which
we try to circumvent with various plausible
explanations in \S~\ref{metallicity}.

The repeated starbursting scenario predicts a number 
of independent observables whose testing is important
but goes beyond the scope of the paper.
They represent a natural extension of the present
work. In order to illustrate the possibilities,
we will outline two examples.
\lsb s must have a stellar population corresponding
to short star formation episodes in between quiescent 
gaps. In principle, one can distinguish between a 
(low level of) continuous star formation 
and a more violent
episodic star formation with a period of 0.3 Gy.  
\citet{leo96} put forward a 
spectroscopic index to detect post-starburst
galaxies. The technique is well suited for determining the 
time elapsed from the last (young) starburst, and it has been calibrated  
by \citet{leo00}. Depending on the noise level,
one can apply the method to selected \lsb\ SDSS spectra,
or to averages of similar \lsb\ spectra.
Another testable prediction of the scenario
has to do with the metallicity
inferred from emission lines in \lsb\ galaxies 
(\S~\ref{metallicity}). It should
overestimates the true galaxy metallicity
and, in particular, the metallicity of the
stellar content. 
Studies of stellar metallicity can be 
carried out using integrated galaxy spectra, provided
that they have enough 
signal-to-noise ratio and spectral resolution
to show absorption lines \citep[e.g.][]{ter90,wor94}.
These two testable predictions provide a flavor
for other tests to come.

%
%





\begin{acknowledgements}
Thanks are due to R. D\'\i az Campos for training with the
SQL queries, to J. Betancort for clarifying
discussions on the LF normalization,
and to A. Vazdekis for pointing out the work 
by \citet{leo96}.
The authors are particularly indebted to
A. D\'\i az,  M.~Mas-Hesse, R.~Terlevich, 
P.~Papaderos,
and
J.~V\'\i lchez for
stimulating discusions during the meetings of  
the {\em estallidos} collaboration.
This work has been partly funded by the 
Spanish {\em Ministerio de Educaci\'on y Ciencia}, proyect
AYA~2007~67965.
We thank the referee for helping us to clarify
some of the arguments in the paper.
%
%
Funding for the SDSS and SDSS-II has been provided by the 
Alfred P. Sloan Foundation, the Participating Institutions, 
the National Science Foundation, the U.S. Department of Energy, 
the National Aeronautics and Space Administration, 
the Japanese Monbukagakusho, the Max Planck Society, and the Higher Education 
Funding Council for England. T
he SDSS Web Site is http://www.sdss.org/.
    The SDSS is managed by the Astrophysical Research Consortium for the 
Participating Institutions. The Participating Institutions are the 
American Museum of Natural History, Astrophysical Institute Potsdam, 
University of Basel, University of Cambridge, Case Western Reserve University, 
University of Chicago, Drexel University, Fermilab, 
the Institute for Advanced Study, the Japan Participation Group, 
Johns Hopkins University, the Joint Institute for Nuclear Astrophysics, 
the Kavli Institute for Particle Astrophysics and Cosmology, 
the Korean Scientist Group, the Chinese Academy of Sciences (LAMOST), 
Los Alamos National Laboratory, the Max-Planck-Institute for 
Astronomy (MPIA), the Max-Planck-Institute for Astrophysics (MPA), 
New Mexico State University, Ohio State University, 
University of Pittsburgh, 
University of Portsmouth, Princeton University, 
the United States Naval Observatory, 
and the University of Washington.
\end{acknowledgements}


\appendix
\section{Luminosity function estimate}\label{appa}

The LF, $\Phi(M)$, is  defined as the 
number of galaxies with absolute magnitude $M$ per unit 
of magnitude and  unit of volume. 
We use it to describe the local universe and, 
therefore, $\Phi(M)$ is not expected to change with  
the position in space.  
The number of galaxies per unit  volume, 
$n_0$, is just,  
\begin{equation}
n_0=\int_{M_l}^{M_u}
\Phi(M)\,dM,
\label{this_is_needed}
\end{equation}
where only absolute magnitudes in between the limits 
$M_l \leq M \leq M_u$ are considered. 
Using the definition of LF, one can easily write down 
the number of galaxies to be expected 
in an apparent magnitude limitted catalog, namely,
\begin{equation}
N=\int_{M_l}^{M_u}
\Phi(M)\,V_{\rm max}(M)\,dM.
\label{this_is_needed_too}
\end{equation}
As usual, the symbol $V_{\rm max}(M)$ stands 
for the volume of universe covered by the catalog 
where galaxies of absolute magnitude $M$ 
have apparent magnitudes smaller than the catalog 
threshold \citep[e.g.,][]{tak00}.

Maximum likelyhood estimates of LFs are favored
in the current literature \citep[e.g.,][]{lin96,bla01}. 
They were pioneered 
by \citet{efs88} and, among the quoted advantages,
they present a number of desirable asymptotic error 
properties\footnote{They are consistent, i.e., they tend 
to the parameter to be estimated when the sample increases, and
the distribution becomes a normal of minimum variance 
for large samples \citep[see, e.g.,][\S~7.2 ]{mar71}.}.
Note, however, that these methods are not unbiased (the expected
value of the estimate is not necessarily the parameters to be estimated).
The method is based on the following principle;
given the redshift $\zeta$ of an observed galaxy,
the probability that it has an absolute magnitude $M$ is, 
\begin{equation}
P(M\,|\,\zeta)=\Phi(M)\Big/\int^{M_{\rm max}(\zeta)}_{M_{\rm min}(\zeta)}\Phi(M')\,dM',
\end{equation}
where the normalization gives all the galaxies that we
are allowed to observe at $\zeta$. Since our catalog is limited
in apparent magnitude $m_{\rm max}$, 
\begin{equation}
m \leq m_{\rm max},
\end{equation}
only galaxies bright
enough would be observable at this redshift.
Considering the definition of distance 
modulus,
\begin{equation}
DM(\zeta)=m-M,
\label{dm}
\end{equation}
then, 
\begin{equation}
M \leq M_{\rm max}(\zeta)=m_{\rm max}-DM(\zeta).
\label{mmax}
\end{equation}
Similarly, the sample would have a minimum
apparent magnitude $m_{\rm min}$ (e.g., given by the 
brightest  galaxy in the sample), which sets the 
minimum magnitude to be observed,
\begin{equation}
M \geq M_{\rm min}(\zeta)=m_{\rm min}-DM(\zeta).
\label{mmin}
\end{equation}
Assuming that the probability of observing each galaxy
is independent from the rest of galaxies,
the likelyhood function $\mathcal{L}$ is just the
product of the probability of observing each galaxy,
\begin{equation}
\mathcal{L}=\Pi_j\,P(M_j\,|\,\zeta_j),
\end{equation}
where the index $j$ spans from 1 to $N$. 
The maximum likelyhood 
estimate maximizes $\mathcal{L}$, which is
equivalent to maximizing its logarithm,
\begin{equation}
\log\mathcal{L}=\sum_j\log\Phi(M_j)
-\sum_j\log\Big[\int^{M_{\rm max}(\zeta_j)}_{M_{\rm min}(\zeta_j)}\Phi(M')\,dM'\Big].
\label{loglike}
\end{equation}
The next step consist in parameterizing the LF in terms of a number $K$
of free parameters $X_i$,
\begin{equation}
\Phi(M)=\phi(M,X_i),
\end{equation}
with $i=1,\dots K$. 
Various representations $\phi(M,X_i)$ 
can be found in the literature, e.g.,
a stepwise function \citep{efs88},
a Schechter function \citep{lin96},
a collection of Gaussians \citep{bla03b}.
We choose yet another representation, namely,
natural cubic splines. The reasons are (1) the LF is
simple and fast to compute, (2) the integral of the LF is
also fast to compute since it follow 
directly from the spline interpolation,
and (3) it automatically provides a smooth $\Phi(M)$.
Speed is always an appealing feature
since the maximization of $\log\mathcal{L}$ is carried 
out
iteratively, and requires many evaluations 
of the LF.

The normalization  of the LF is not constrained 
by the likelyhood function, which is independent 
of a global scaling factor (equation~[\ref{loglike}]). 
Another complementary method is required to estimate
the number density of galaxies $n_0$. Such method is
often the minimum variance estimate by \citet{dav82}.
We use a simple version of such estimate, where 
all the galaxies in the sample are equally 
weighted, thus avoiding assuming a particular
covariance of the galaxy catalog. 
It corresponds to the so-called $n_3$ in the original 
paper by \citet{dav82}, and it has been used elsewhere 
\citep[e.g.,][]{bor02}. 
The expression can be derived by
combining equations~(\ref{this_is_needed}) and 
(\ref{this_is_needed_too}),
which yield,
\begin{equation}
n_0=N\Big/\int_{M_l}^{M_u}
\varphi(M)\,V_{\rm max}(M)\,dM.
\label{myn0}
\end{equation}
All items in the right-hand-side of 
the previous expression are known;
the normalized LF $\varphi(M)$, 
\begin{equation}
\varphi(M)=\Phi(M)\Big/\int_{M_l}^{M_u}\Phi(M')\,dM',
\end{equation}
is provided by the maximum likelyhood procedure, whereas
$V_{\rm max}(M)$ follows from the magnitude limit and the solid
angle of the catalog \citep[e.g.,][]{tak00}.

The actual maximization of the likelyhood in
equation~(\ref{loglike}) is carried out using 
the standard Powell method \citep[][]{pre88}. 
Error bars are assigned by bootstrapping \citep[e.g.,][]{moo03,bla03b},
where one constructs bootstrap re-samples 
by choosing at random galaxies from the original data set.
Then the application of the LF retrieval procedure
to all bootstrap re-samples yields a set
of LFs with the spread of values to be expected from
the true error distribution \citep[][]{moo03}. 
We use the standard deviation of such bootstrap 
distribution as our error bars. 

\begin{figure}
   \centering
   \includegraphics[width=0.5\textwidth]{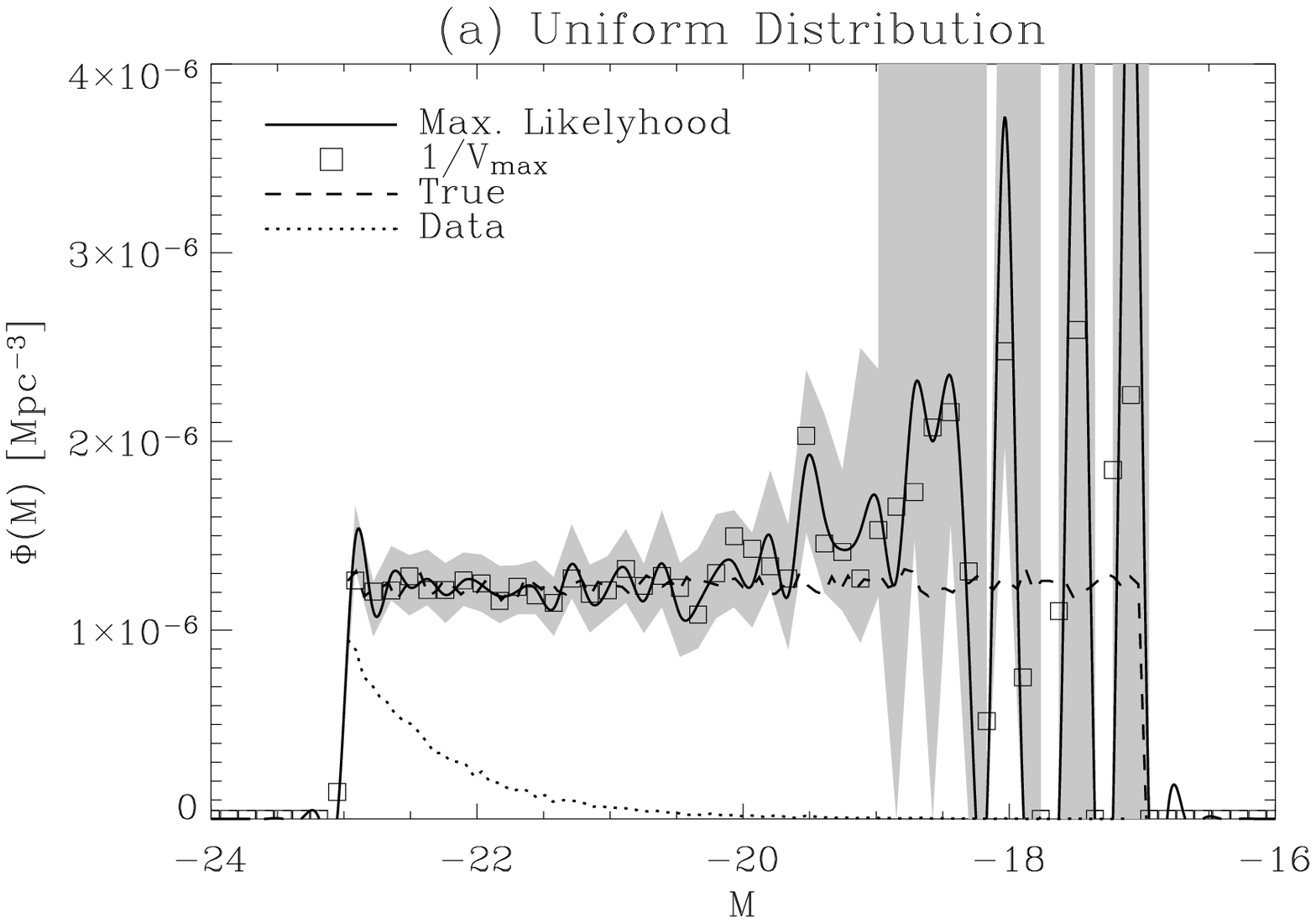}
   \includegraphics[width=0.5\textwidth]{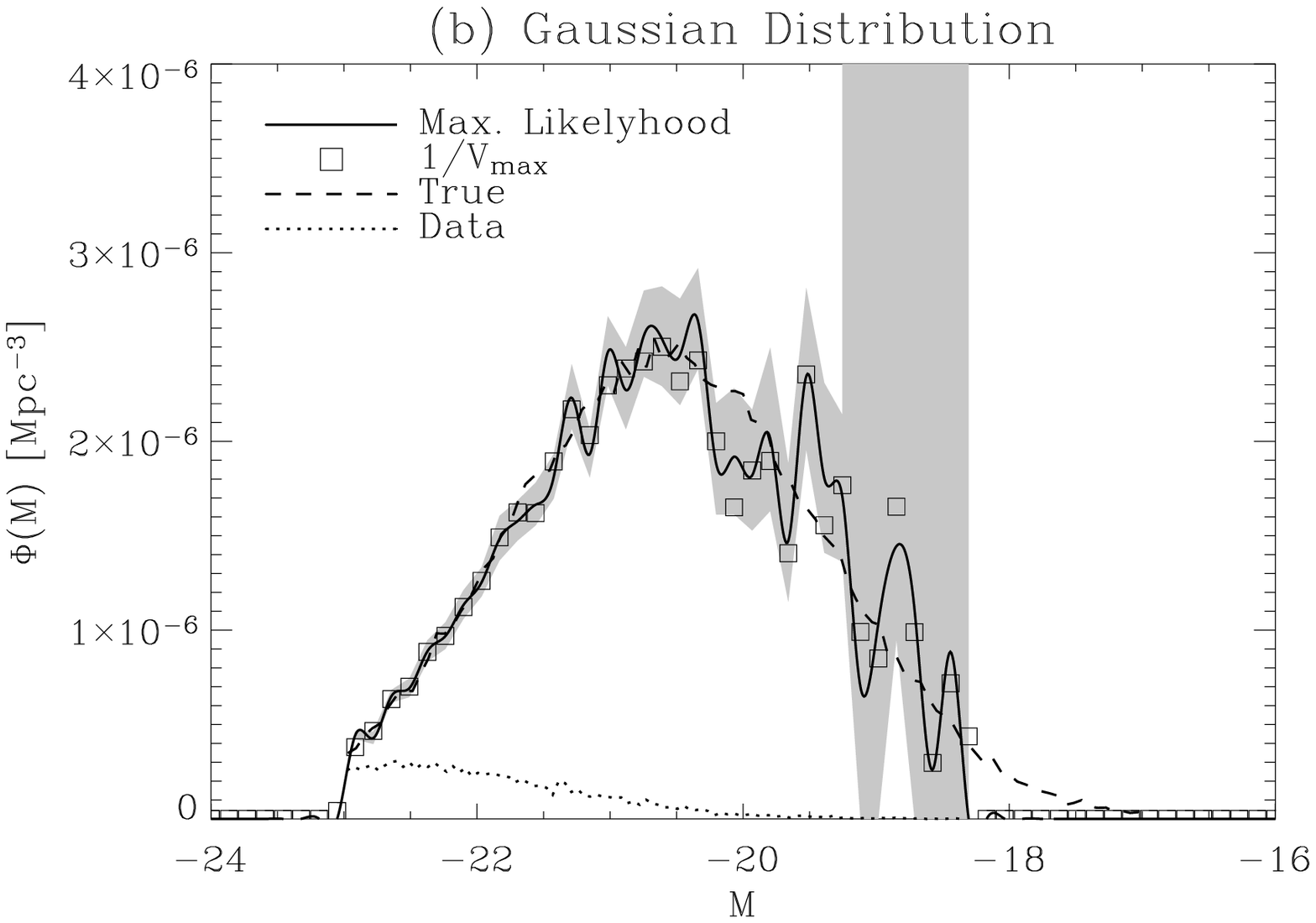}
     \caption{Numerical tests to our maximum likelyhood LF estimate.
	The various types of lines and symbols correspond to the
	true LF (the dashed line), the restored $1/V_{\rm max}$ LF
	(square symbols), and 
	the maximum likelyhood LF (the solid line, with the shaded
	region representing the error bars). The dotted line
	corresponds to the original data offered to the LF retrieval
	routine. (a) Case where the true LF is a uniform distribution
	between magnitudes -23 and -17. (b) Same as (a) except that the
	underlying LF is a truncated Gaussian.
              }
         \label{tests_fig}
   \end{figure}

Numerical tests have been carried out to check the procedure.
We choose a large number of galaxies ($10^{5}$--$10^{6}$) with 
random absolute magnitude according to Gaussian or 
uniform distributions.  These galaxies are randomly uniformly 
spread in space within a sphere of radius $\zeta=0.35$.  
Apparent magnitudes are computed using these redshifts, 
and galaxies fainter than the assumed catalog cutoff are drooped 
from the sample (we take $m_{\rm max}=17.7$). 
This biased sample is then used to fed the procedure.
Two examples of true and 
restored LFs are given in Fig.~\ref{tests_fig}.
The solid lines correspond to our maximum likelyhood
estimate, whereas the dashed lines show the {\em true} 
histogram derived from the synthetic data before the 
apparent magnitude threshold  is introduced. The two 
of them agree within error 
bars, which are given as shaded areas. Moreover, the maximun likelyhood 
estimates
also agree with the $1/V_{\rm max}$ estimates included in 
the same plot 
(see, e.g.,  \citealt{tak00} for a description of traditional
$1/V_{\rm max}$ method by \citealt{sch68}). The agreement is not 
specific of these particular realizations but is a general 
property, and it seems to hold both independently of the original 
distribution, and the number of galaxies in the sample.
In order to illustrate the magnitude of the correction
carried out by the LF retrieval routine, 
Fig.~\ref{tests_fig} also includes the distributions of magnitudes
of the some $10^{4}$ galaxies  given to the program (the dotted lines).

\bibliographystyle{aa}

 \end{document}